\documentclass[aps,prl,reprint,twocolumn,superscriptaddress,floatfix,nofootinbib,longbibliography]{revtex4-1}
\usepackage{graphicx,amsmath,amsfonts,amssymb,amsthm,xr}
\usepackage{epsfig,amsmath,amssymb,color,dsfont,upgreek,physics}
\usepackage{mathrsfs}
\usepackage{mathtools}
\usepackage{bbold}
\usepackage{comment}
\usepackage{float}

\usepackage[bookmarks=true,colorlinks,linkcolor=OrangeRed,urlcolor=NavyBlue,citecolor=RoyalBlue]{hyperref}
\usepackage[dvipsnames]{xcolor}
\usepackage{orcidlink}

\definecolor{mygold}{rgb}{0.93,0.69,0.13}

\definecolor{mypurple}{rgb}{0.49,0.18,0.56}

\newcommand{\beq}{\begin{equation}}
\newcommand{\eeq}{\end{equation}}

\newcommand{\eps}{\varepsilon}

\begin{document}
\title{Unifying Finite-Temperature Dynamical and Excited-State Quantum Phase Transitions}
\author{\'{A}ngel L.~Corps${}^{\orcidlink{0000-0002-0666-6642}}$}
\email{corps.angel.l@gmail.com}
\affiliation{Instituto de Estructura de la Materia, IEM-CSIC, Serrano 123, E-28006 Madrid, Spain}
\affiliation{Grupo Interdisciplinar de Sistemas Complejos (GISC), Universidad Complutense de Madrid, Av.~Complutense s/n, E-28040 Madrid, Spain}
\author{Armando Rela\~{n}o${}^{\orcidlink{0000-0002-1670-1544}}$}
\email{armando.relano@fis.ucm.es}
\affiliation{Grupo Interdisciplinar de Sistemas Complejos (GISC), Universidad Complutense de Madrid, Av.~Complutense s/n, E-28040 Madrid, Spain}
\affiliation{Departamento de Estructura de la Materia, F\'{i}sica T\'{e}rmica y Electr\'{o}nica,
Universidad Complutense de Madrid, Av.~Complutense s/n, E-28040 Madrid, Spain}
\author{Jad C.~Halimeh${}^{\orcidlink{0000-0002-0659-7990}}$}
\email{jad.halimeh@physik.lmu.de}
\affiliation{Department of Physics and Arnold Sommerfeld Center for Theoretical Physics (ASC), Ludwig-Maximilians-Universit\"at M\"unchen, Theresienstra\ss e 37, D-80333 M\"unchen, Germany}
\affiliation{Munich Center for Quantum Science and Technology (MCQST), Schellingstra\ss e 4, D-80799 M\"unchen, Germany}
\affiliation{Dahlem Center for Complex Quantum Systems, Freie Universit\"at Berlin, 14195 Berlin, Germany}
\begin{abstract}
In recent years, various notions of dynamical phase transitions have emerged to describe far-from-equilibrium criticality. A unifying framework connecting these different concepts is still missing, and would provide significant progress towards understanding far-from-equilibrium quantum many-body universality. Initializing our system in a thermal ensemble and subsequently performing quantum quenches in the Lipkin-Meshkov-Glick model, we establish a direct connection between excited-state quantum phase transitions (ESQPTs) and two major types of dynamical phase transitions (DPTs), by relating the phases of the latter to the critical energies and conservation laws in the former. Our work provides further insight into how various concepts of non-ground-state criticality are intimately connected, paving the way for a unified framework of far-from-equilibrium universality.
\end{abstract}

\date{\today}
\maketitle

\textbf{\textit{Introduction.---}}Phase transitions and critical phenomena, along with the resulting features of universality and scaling, are well-understood concepts in equilibrium. In a far-from-equilibrium setting, a unified framework of these notions is still missing. The pursuit of an overarching theory of far-from-equilibrium quantum many-body criticality has recently led to different concepts of nonequilibrium phase transitions \cite{Mori_review,Heyl_review}. The first one is related to the dynamics of the equilibrium Landau order parameter, which is connected to the spontaneous breaking of a global symmetry in the ground state \cite{Cardy_book}. Upon quenching a given symmetry-broken initial state, if the long-time steady state exhibits a nonzero (zero) order parameter, then the system is in a symmetry-broken (symmetry-preserved) \textit{dynamical} phase \cite{Moeckel2008,Eckstein2008}. The value of the quench parameter separating these two phases is the \textit{dynamical quantum critical point}. This type of dynamical phase transition has been dubbed DPT-I, and has been studied in various systems, including mean-field models \cite{Sciolla2010,Sciolla2011,Zunkovic2016,Homrighausen2017anomalous,Lang2018concurrence,Lang2018geometric,Corps2022dynamical,Corps2023theory,Corps2023relaxation}, the Hubbard model \cite{Eckstein2009,Moeckel2010,Tsuji2013}, $O(N)$ model \cite{Chandan2013,Maraga2015,Smacchia2015,Chiocchetta2017,Halimeh2021}, long-range quantum spin chains \cite{Halimeh2017,Zunkovic2018dynamical}, among others \cite{Gambassi2011,Langen2016,Marcuzzi2016}.

Another approach to dynamical criticality encompasses the construction of a dynamical analog of the thermal free energy. This becomes straightforward when recognizing the overlap of the time-evolved wave function with the initial state as a boundary partition function where evolution time stands for a complex inverse temperature \cite{Heyl2013,Heyl2014,Heyl2015}. By taking the negative of the logarithm of this overlap in the thermodynamic limit, one obtains the return rate, which is the sought-after dynamical analog of the thermal free energy. Nonanalyticities in the return rate are thus \textit{dynamical quantum phase transitions} (DQPTs) at \textit{critical evolution times}. DQPTs are also referred to as DPT-II, and have been extensively studied in nonintegrable short-range quantum spin systems \cite{Karrasch2013dynamical,Vajna2014disentangling,Andraschko2014dynamical}, long-range quantum many-body models \cite{Zunkovic2016,Halimeh2017dynamical,Zauner-Stauber2017probing,Homrighausen2017anomalous,Zunkovic2018dynamical,Defenu2019dynamical,Uhrich2020out,Halimeh2021dynamical,Corps2022dynamical,Corps2023theory,Deutsch2023macrostates,Corps2023mechanism}, topological systems \cite{Vajna2015topological,Schmitt2015dynamical,Sedlmayr2018bulk,Hagymasi2019dynamical,Maslowski2020quasiperiodic,Porta2020topological,Okugawa2021mirror,Sedlmayr2022dynamical,Maslowski2023dynamical}, higher-dimensional models \cite{Schmitt2015dynamical,Bhattacharya2017emergent,Weidinger2017dynamical,Heyl2018detecting,DeNicola2019stochastic,Hashizume2022dynamical,Hashizume2020hybrid,kosior2024vortex}, systems initialized in thermal ensembles \cite{Abeling2016quantum,Bhattacharya2017mixed,Heyl2017dynamical,Sedlmayr2018fate,Lang2018concurrence,Lang2018geometric}, high-energy models known as lattice gauge theories \cite{Zache2019,Huang2019dynamical,Pedersen2021,Jensen2022,Halimeh2021achieving,VanDamme2022dynamical,Mueller2023quantum,Pomarico2023dynamical,VanDamme2023Anatomy,Osborne2023probing}, non-Hermitian systems \cite{Zhou2018Dynamical,Zhou2021Non-Hermitian,Hamazaki2021Exceptional,Nag2022Anomaly,Nag2023Finite,mondal2024persistent}, short-range interacting systems with broken time-translation symmetry \cite{Kosior2018a,Kosior2018b}, and disordered models \cite{Halimeh2019dynamical,Trapin2021}. Furthermore, they have been the subject of several successful experiments \cite{Jurcevic2017,Flaeschner2018,Nie2020experimental}.

A different source of criticality beyond the ground state is given by excited-state quantum phase transitions (ESQPTs) \cite{Caprio2008,Cejnar2021}. They consist in a generalization of quantum phase transitions to excited states, typically manifested as a singularity of the density of states and the level flow. Notwithstanding, the main consequences are dynamical, like huge decoherence \cite{Relano2008,PerezFernandez2009}, singularities in quench dynamics \cite{PerezFernandez2011,Santos2015,Lobez2016,PerezBernal2017,Kloc2018}, feedback control in dissipative systems \cite{Kopylov2015}, quantum work statistics \cite{Wang2017}, symmetry-breaking equilibrium states \cite{Puebla2013,Puebla2014}, dynamical instabilities \cite{Bastidas2014}, irreversibility without energy dissipation \cite{Puebla2015}, and reversible quantum information spreading \cite{Hummel2019}. It has been recently shown that they may give rise to a phase diagram composed by different dynamical phases, each one characterized by a set of (generally noncommuting) constants of motion \cite{Corps2021,Corps2022,Corps2023}. 

A pertinent question is how these different concepts of nonequilibrium quantum phase transitions are related to one another. When it comes to DPT-I and DPT-II, a connection has been established between the phase of the former and the type of the latter \cite{Homrighausen2017anomalous}: Starting in a symmetry-broken initial state, if the long-time steady state breaks (preserves) the global symmetry of the quench Hamiltonian, then the DPT-II will be of the anomalous (regular) type. This connection also persists at finite temperature \cite{Lang2018concurrence,Lang2018geometric}. This potentially allows drawing connections between far-from-equilibrium critical exponents arising in both these DPTs \cite{Halimeh2021Local,Wu2020}. Nevertheless, the connection between ESQPTs and DPTs, in particular at finite temperature, is still ambiguous. Given the potential of understanding DPT criticality from that of ESQPTs, it is therefore important to investigate if a direct connection exists. This is the purpose of this Letter.

\textbf{\textit{Model.---}} Although our arguments are general, we chose a collective model, allowing us to reach large system sizes, as an illustration. It is the transverse-field Ising model with infinite-range interactions, which coincides with a version of the Lipkin-Meshkov-Glick (LMG) Hamiltonian \cite{Lipkin1965,Meshkov1965,Glick1965},
\beq\label{eq:ham}
\hat{H}=-\frac{\lambda}{N}\hat{J}_{x}^{2}+h\hat{J}_{z}
\eeq
The total collective spin operator commutes with the Hamiltonian, $[\hat{H},\mathbf{J}^{2}]=0$, allowing us to separate spin sectors labeled by the eigenvalues of $\mathbf{J}^{2}$, $j(j+1)$, the dimension of each being $D(j)=2j+1$. Hamiltonian~\eqref{eq:ham} also has a discrete $\mathbb{Z}_2$ symmetry generated by a $\pi$-rotation around the $z$-axis. In each $j$-sector it is represented by a parity operator, $\hat{\Pi}=\textrm{e}^{i \pi (\hat{J}_z + j)}$, which allows to classify the Hamiltonian eigenstates according to $\hat{\Pi}\ket{E_{n,\pm}}=\pm\ket{E_{n,\pm}}$, $n=0,1,\ldots$

The full model displays two critical phenomena. At $\lambda_c=\lambda$ there is a quantum phase transition (QPT): for $\lambda>\lambda_c$ the ground state is ferromagnetic and the symmetry generated by $\hat{\Pi}$ is broken; for $\lambda<\lambda_c$ the ground state is symmetric. In the first case, there exists also a thermal phase transition with a critical inverse temperature given by $\beta_c = 2h^{-1} \textrm{arctanh} (h/\lambda)$; at lower temperatures, the system is ferromagnetic, and the $\mathbb{Z}_2$ symmetry is broken.

To make a connection between these facts and ESQPTs and DPTs, we work with all $j$-sectors, $j=0,1,...,N/2$, each with a degeneracy factor of $g(N,j)=\frac{1+2j}{1+j+N/2}\binom{N}{N/2-j}$, so $\sum_{j=0}^{N/2}g(N,j)D(j)=2^{N}$. 
As noted in \cite{PerezFernandez2017dicke} for a similar fully connected model, each $j$-sector is completely independent of the others, and therefore it can be described by Hamiltonian~\eqref{eq:ham} with an effective coupling constant given by $\lambda_{\textrm{eff}}=2 j \lambda /N$.
This means that each $j$-sector has its own critical points. For the QPT, it is $\lambda_c (j)=hN/(2j)$.
Thus, above the critical point for the global QPT, $\lambda_c=h$,
some of the $j$-sectors are in the ferromagnetic ground-state phase, and some others are in the paramagnetic ground-state phase; they are separated a {\em critical} value $j_c(\lambda)=Nh/(2\lambda)$.
If $j>j_c$, the corresponding sector is in the ferromagnetic phase, and the opposite occurs if $j<j_c$. This argument is important to understand the behavior of ESQPTs: 

(i) If $\lambda<\lambda_c$, all the $j$-sectors are in the paramagnetic phase. Therefore, there are no critical energies and their ground-state energies are $\varepsilon_{\textrm{GS}} (j) = - 2 h j/N$, where $\varepsilon=2 E /N$.

(ii) If $\lambda>\lambda_c$, the behavior is more involved:

(a) If $j>j_c(\lambda)$, the $j$-sector is in the ferromagnetic ground-state phase. Therefore, it has a critical energy below which all its energy levels are pairwise degenerate in the infinite-size limit. This is the ESQPT energy:
\beq
\label{eq:critj}
\varepsilon_c (j) = - \frac{2h j}{N}.
\eeq
The corresponding ground-state energy is
\beq
\label{eq:gsj}
\varepsilon_{\textrm{GS}} (j) = -\left[ 2 \lambda \left( \frac{j}{N} \right)^2 + \frac{h^2}{2 \lambda} \right].
\eeq
Note that Eqs.~\eqref{eq:critj} and \eqref{eq:gsj} coincide if $j=j_c$, $\varepsilon_c(j_c)=-h^2/\lambda$.

(b) If $j<j_c$, the sector is in the paramagnetic phase. Therefore, none of its eigenlevels are degenerate, with a ground-state energy of
$\varepsilon_{\textrm{GS}} (j) = - 2 h j/N$. 

The main consequence of these facts is that we can define {\em two} critical energies for the full Hamiltonian, $\varepsilon_{c1}=-h$ and $\varepsilon_{c2}=-h^2/\lambda>\varepsilon_{c1}$. If $\varepsilon < \varepsilon_{c1}$, all the energy levels are degenerate in pairs; and if $\varepsilon_{c1} \leq \varepsilon < \varepsilon_{c2}$, pairwise degenerate and non-degenerate energy occur simultaneously.
Above $\varepsilon_{c2}$, there are no degeneracies.

\textbf{\textit{Equivalence of ESQPTs and DPTs.---}}All these features refer to static or equilibrium properties of the LMG Hamiltonian. However, DPTs are nonequilibrium phenomena. To establish a link between them, we focus on a dynamical property of a class of ESQPTs. In Refs.~\cite{Corps2021,Corps2022,Corps2023} it is shown that for a wide class of models to which the LMG belongs, there are two additional constants of motion below the critical energy of the ESQPT related to the order parameter of the QPT and the operator generating the $\mathbb{Z}_2$ symmetry: $\hat{\mathcal{C}}=\textrm{sign}(\hat{J}_x)$ and $\hat{\mathcal{K}}=(i/2) [\hat{\mathcal{C}},\hat{\Pi}]$. Thus, if a nonequilibrium protocol leads the system into an energy region below the ESQPT, then the dynamics is restricted by the conservation of $\hat{\mathcal{C}}$ and $\hat{\mathcal{K}}$. As a consequence, quenching {\em an initial symmetry-breaking state} polarized along the ferromagnetic axis cannot lead the order parameter $\langle \hat{J}_x \rangle$ to change sign in its dynamics. On the contrary, there are no restrictions if the energy is above the ESQPT. These dynamical features are also expected for noncollective models with an equilibrium symmetry-breaking phase \cite{Corps2023,Corps2023arxiv}.

From these facts, we propose the main conclusion of this Letter: {\em there are only two possible dynamical phases (DPs) starting from an equilibrium symmetry-breaking initial state}:

{\bf DPa.} A constant value for $\hat{\mathcal{C}}$ and $\hat{\mathcal{K}}$, together with the order parameter $\langle \hat{J}_x \rangle$ oscillating around a nonzero value, without changing sign.

{\bf DPb.} Both $\hat{\mathcal{C}}$, $\hat{\mathcal{K}}$, and $\langle \hat{J}_x \rangle$ oscillating around zero.
These two dynamical phases are separated by the critical energy of the ESQPT.

\begin{figure}[h!]
\hspace{-1.4cm}\includegraphics[width=0.55\textwidth]{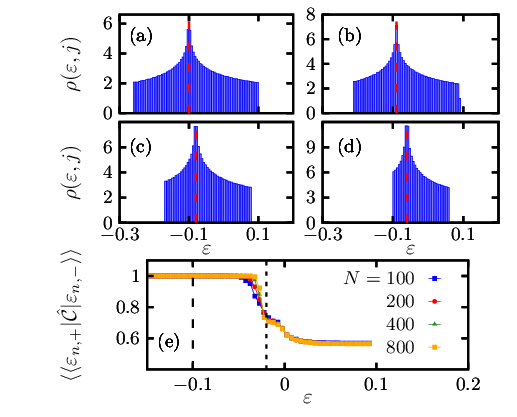} 

\caption{(a-d) Density of states for the LMG model with $N=10000$ particles and parameters $h=0.1$, $\lambda=0.5$. Each panel corresponds to the level density as obtained for a given $j$-sector. (a) $j=N/2=5000$, (b) $j=4500$, (c) $j=4000$, (d) $j=3000$.  The eigenvalues are rescaled as $\eps=2E/N$. Black vertical lines mark the ESQPT critical energy of each $j$-sector \eqref{eq:critj}. (e) Expectation value of $\hat{\mathcal{C}}$ in states of different parity as a function of energy. Model parameters are $h=0.1$, $\lambda=0.5$, with $N$ as indicated in the panel. Dashed vertical lines mark the energies $\eps=-h=-0.1$ and $\eps=-h^{2}/\lambda=-0.02$. Points represent an average over a small energy window containing all $j$-sectors, and with the corresponding degeneracy factors $g(N,j)$ appropriately taken into account.}
\label{fig:dos}
\end{figure}

To illustrate this picture, we show in Fig.~\ref{fig:dos} the density of states for $\lambda=0.5$, $h=0.1$, and different values of $j$. All of them verify $j>j_c(\lambda)$, and therefore exhibit ESQPTs. It is clearly seen that the corresponding critical energy, identified by the logarithmic divergence in the density of states, shifts to lower energies as $j$ is increased. This means that the populated sectors of $j$ play a fundamental role in the dynamics. Let us suppose that we prepare a state with $\epsilon=-0.07$. This value is above the critical energy of panels (a)-(c), and below the one on panel (d). If the expectation value of $\mathbf{J}^2$ in our state, which is a conserved quantity, is narrowly picked around $j/N>0.35$, then neither $\hat{\mathcal{C}}$ nor $\hat{\mathcal{K}}$ are constants, and therefore the order parameter $\langle J_x \rangle$ must oscillate around zero. On the contrary, from Eqs.~\eqref{eq:critj} and \eqref{eq:gsj}, we can conclude that if $\sqrt{0.06}<j/N<0.35$, then both $\hat{\mathcal{C}}$ and $\hat{\mathcal{K}}$ are constant, and therefore $\langle \hat{J}_x \rangle$ cannot cross $\langle \hat{J}_x \rangle=0$.

In Fig.~\ref{fig:dos}(e) we represent the consequences of the previous facts for the full Hamiltonian. As discussed in \cite{Corps2021}, the constancy of $\hat{\mathcal{C}}$ requires that $\bra{\varepsilon_{n,-}} \hat{\mathcal{C}} \ket{\varepsilon_{n,+}}= \pm 1$ for degenerate energy levels $\varepsilon_{n,-}=\varepsilon_{n,+}$. The results indicate that this is globally fulfilled if $\varepsilon < \varepsilon_{c2}$ in the thermodynamic limit. The reason is that, the lower the value of $j$, the larger the degeneracy factor, $g(N,j)$; therefore, only the lowest possible value of $j$, which gives rise to the highest critical energy $\varepsilon_c(j)$, contributes to the dynamics of $\hat{\mathcal{C}}$ in the thermodynamic limit. Therefore, we can expect symmetry-broken {\em thermal} states if $\varepsilon<\varepsilon_{c2}$, whose associated temperature is below the critical temperature of the phase transition, $T<T_{c}$. However, this is not enough to determine the dynamics of a thermal state subjected to a nonequilibrium process. As $\hat{\mathbf{J}}^2$ is conserved, the population of each $j$-sector must be taken into account to determine whether the final state is above or below the critical energies of the corresponding ESQPTs. It is worth to note that the same qualitative result shown in Fig.~\ref{fig:dos}(e) has been observed in the transverse-field Ising model with long-range interactions \cite{Corps2023}, in which $\hat{\mathbf{J}}^2$ is not conserved. Hence, the same classification in two dynamical phases is expected for noncollective models.

To test our hypothesis, we have performed a set of numerical experiments on the LMG model. In all of them, we prepare an initial state in the ferromagnetic phase, with $\varepsilon<\varepsilon_{c1}$. As $\hat{\mathbf{J}}^2$, $\hat{\Pi}$, $\hat{\mathcal{C}}$, $\hat{\mathcal{K}}$ are conserved under these circumstances, the most general equilibrium state is
\begin{equation}\label{eq:initialstate}
\hat{\rho} = \frac{1}{Z} \textrm{e}^{-\beta \hat{H} - \mu_c \hat{\mathcal{C}}- \mu_k \hat{\mathcal{K}}- \mu_{\pi} \hat{\Pi} -\mu_j \hat{\mathbf{J}}^2},
\end{equation}
where $Z$ is the partition function ensuring that $\textrm{Tr}[\hat{\rho}]=1$, and $\mu_c$, $\mu_k$, $\mu_{\pi}$, $\mu_{j}\in\mathbb{R}$ are free parameters linked to the initial values of $\langle \hat{\mathcal{C}} \rangle$, $\langle \hat{\mathcal{K}} \rangle$, $\langle \hat{\Pi} \rangle$, and $\langle \hat{\mathbf{J}}^2 \rangle$.  To study the dynamics, we start from an initial state $\hat{\rho}_{i}$ of the form \eqref{eq:initialstate} with $\mu_k = \mu_{\pi} = \mu_j=0$, $\mu_c=100$ and $\beta=5$, though our conclusions also hold for other values (see Supplemental Material \cite{SM}). The initial Hamiltonian, $\hat{H}_{i}$, has parameters $\lambda=0.5$ and $h_{i}=0$. This choice gives rise to a polarized thermal state, with $\langle \hat{J}_x \rangle < 0$. We then quench the initial state with a final Hamiltonian, $\hat{H}_{f}$, with different $h_{f}=0.1,0.15,0.2,0.3$ and $\lambda=0.5$. The time-evolved density operator at time $t$ is $\hat{\rho}_{f}(t)=e^{-i\hat{H}_{f}t}\hat{\rho}_{i}e^{i\hat{H}_{f}t}$. 
Since $\hat{\mathbf{J}}^{2}$ is conserved by Eq.~\eqref{eq:ham}, the distribution $P(j)$ of populated $j$-sectors remains unchanged in the wake of the quench. The dynamics will be dominated by $j$-sectors with large $P(j)$.

\begin{figure}[h!]
\hspace{-3.4cm}\includegraphics[width=0.65\textwidth]{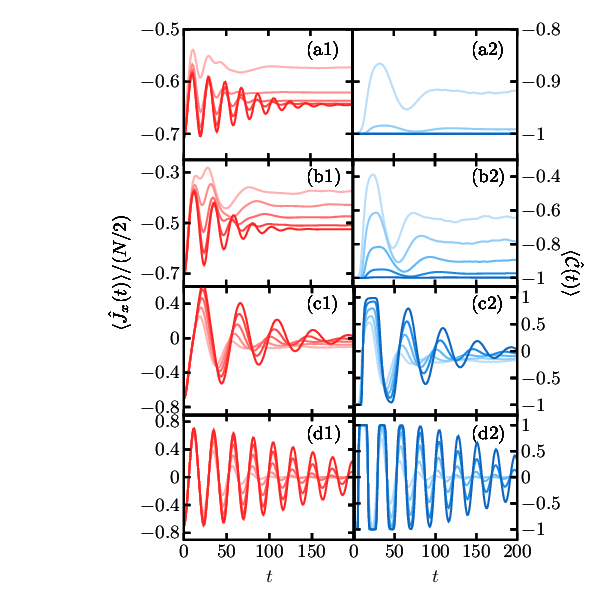}
\caption{Instantaneous values of the magnetization, $\hat{J}_{x}$, (left), and the $\hat{\mathcal{C}}$ operator, (right), following different quenches with $\beta=5$, $\lambda=0.5$ and $h_{i}=0$. (a1,2) $h_{f}=0.1$; (b1,2) $h_{f}=0.15$; (c1,2) $h_{f}=0.2$; (d1,2)  $h_{f}=0.3$. System sizes are $N=100,200,400,800,1600$ from light to dark color curves. In all the cases, $\langle \varepsilon \rangle=-0.126$ and the critical energies for the most populated $j$-sector, given by Eq. \eqref{eq:critj}, are: (a1,2) $\varepsilon_c(j_{\textrm{max}})=-0.0711$, (b1,2) $\varepsilon_c(j_{\textrm{max}}) =-0.107$, (c1,2) $\varepsilon_c(j_{\textrm{max}})=-0.142$, (d1,2) $\varepsilon_c(j_{\textrm{max}})=-0.213$.}
\label{fig:JxtCt}
\end{figure}

Figure~\ref{fig:JxtCt} illustrates the dynamical effects of these quenches. We focus first on the largest system size, $N=1600$. For $h_f=0.1$ and $h_f=0.15$, the average quench energy is below the critical energy of the most-populated $j$-sector. We can see that $\langle \hat{J}_x \rangle$ oscillates around a nonzero value (note that the larger the system, the longer the oscillating behavior remains), and $\langle \hat{\mathcal{C}} \rangle$ is perfectly constant; therefore, the system is in the dynamical phase DPa. On the contrary, for $h_f=0.2$ and $h_f=0.3$ the average quench energy is above the critical one, and the dynamics is consistent with the dynamical phase DPb: both $\langle \hat{J}_x \rangle$ and $\langle \hat{\mathcal{C}} \rangle$ oscillate around zero. It is worth to remark that the critical quench separating regular and anomalous DPTs-II is given by $h_f^c \approx 0.1776$ \cite{Lang2018concurrence}. Therefore, our numerical results show that DPa leads to anomalous DPTs-II, and DPb to regular ones.

Notwithstanding, the picture is not so clear for smaller system sizes. For $h_f=0.15$ and $N=100$, $200$, $400$, and $800$, $\langle \hat{J}_x \rangle$ oscillates around a non-zero value, but $\langle \hat{\mathcal{C}} \rangle$ is clearly not constant. To explain this behavior and to understand what is expected to occur in the thermodynamic limit (TL), we perform a finite-size scaling. Results are given in Table~\ref{table2}. We focus there on two quantities: the energy width, $\sigma_{\varepsilon} = \sqrt{\langle \hat{H}^2 \rangle - \langle \hat{H} \rangle^2}$, and a range of critical energies obtained from the population of the different $j$-sectors (see caption for details). The key point is the overlap between these two intervals. For a non-empty overlap, we expect a mixture of DPa and DPb. For those sectors in which $\langle \varepsilon \rangle_j < \varepsilon_c (j)$, $\langle \hat{\mathcal{C}} \rangle_j$ is constant, and $\langle \hat{J}_x \rangle_j$ oscillates around a nonzero value ($\langle \bullet \rangle_j$ stands for the expectation value of an observable in the projection of the state onto the eigenspace in which $\mathbf{\hat{J}}^2$ is equal to $j(j+1)$). And for those sectors in which $\langle \varepsilon \rangle_j > \varepsilon_c (j)$, both $\langle \hat{\mathcal{C}}\rangle_j$ and $\langle \hat{J}_x \rangle_j$ oscillate around zero. Therefore, when putting together all the $j$-sectors, we obtain the intermediate picture observed for $h_f=0.15$ and $N=100$, $200$, $400$, and $800$. On the contrary, if the intervals do not overlap, the dynamical phases are either DPa or DPb.

To extrapolate these result to the TL, we study how the energy width and the range of critical energies change with  system size. A least-squares fit of the data shown in Table~\ref{table2} provides $\sigma_{\varepsilon} \propto N^{-0.516(5)}$ and $\left| \varepsilon_{c,\textrm{max}} - \varepsilon_{c, \textrm{min}} \right| \propto N^{-0.54(2)}$. This means that the only possible nonequilibrium dynamics in the TL is either DPa or DPb, and that an initial state gives rise to either one or the other depending on whether its average quench energy is below or above the critical energy of the ESQPT.

\begin{table}[h]
\begin{center}
 \begin{tabular}{||c c c||} 
 \hline
    $N$ \hspace{0.1cm} & $\langle \varepsilon\rangle\pm \sigma_{\varepsilon} $ \hspace{0.1cm} & $[\varepsilon_{c,\min}, \varepsilon_{c,\max}]$ \\ [0.5ex] 
 \hline\hline
 $100$ \hspace{0.1cm} & $-0.120\pm 0.041$ \hspace{0.1cm} & $[-0.135,-0.057]$ \\[1ex]
 \hline
  $200$ \hspace{0.1cm} & $-0.124\pm 0.028$ \hspace{0.1cm} & $[-0.1275,-0.078]$ \\[1ex]
  \hline
  $400$ \hspace{0.1cm} & $-0.125\pm 0.020$ \hspace{0.1cm} & $[-0.1223,-0.08775]$ \\[1ex]
 \hline
$800$ \hspace{0.1cm} & $-0.126\pm 0.014$ \hspace{0.1cm} & $[-0.1178,-0.09375]$ \\[1ex]
 \hline
 $1600$ \hspace{0.1cm} & $-0.1259\pm 0.0097$ \hspace{0.1cm} & $[-0.1146,-0.09769]$ \\[1ex]
 \hline

\end{tabular}
\end{center}
\caption{Average energy, $\langle\varepsilon\rangle$, width $\sigma_{\varepsilon}$ and estimated maximal and minimal critical energies corresponding to the minimum and maximum $j$-sectors, respectively, with a cumulative probability of 95\%, as a function of the system size $N$ for the quench $h_{i}=0\to h_{f}=0.15$, $\lambda=0.5$.}
\label{table2}
\end{table}

\textbf{\textit{Discussion and outlook.---}}Through analytic arguments and numerical simulations, we have shown that ESQPTs and two major types of DPTs have a direct connection to each other in the LMG model. When the quench energy is below (above) the ESQPT critical points, the long-time steady-state falls in the ferromagnetic (paramagnetic) phase of DPT-I. Given that DPT-I and DPT-II have been shown to be directly connected to each other in the LMG model \cite{Homrighausen2017anomalous,Lang2018concurrence}, this means that ESQPTs are also directly connected to DPT-II.

Demonstrating such a direct connection between ESQPTs and DPTs provides evidence that varying concepts of criticality beyond that of the ground state are may be intimately related. This is promising in the pursuit of an overarching framework for far-from-equilibrium quantum many-body universality.

Our conclusions should be valid in other mean-field models where large enough system sizes are accessible in order to faithfully probe criticality. An interesting question is whether our findings also hold for nonintegrable models where access to the full spectrum is only possible for small system sizes that cannot reasonably discern criticality. This makes it hard to adequately study ESQPTs in such systems, although a direct connection between DPT-I and DPT-II is well-established in them \cite{Halimeh2020quasiparticle}.

Another interesting venue for future work entails connecting the critical exponents extracted from ESQPTs and DPTs. For example, it is known that DPT-I and DPT-II have seemingly disparate critical exponents, but since both DPTs have been shown to coincide \cite{Lang2018concurrence,Halimeh2020quasiparticle}, it is likely that their critical exponents have a direct relation.

\begin{acknowledgements}
\textbf{\textit{Acknowledgements.---}}A.L.C.~and A.R.~acknowledge financial support by the Spanish grants PGC-2018-094180-B-I00,  PID2019-106820RB-C21 and PID2022-136285NB-C31, funded by Ministerio de Ciencia e Innovaci\'{o}n/Agencia Estatal de Investigaci\'{o}n MCIN/AEI/10.13039/501100011033 and FEDER "A Way of Making Europe". A.L.C.~acknowledges financial support from `la Caixa' Foundation (ID 100010434) through the fellowship LCF/BQ/DR21/11880024. J.C.H.~acknowledges financial support through the Emmy Noether Programme of the German Research Foundation (DFG) under grant no.~HA 8206/1-1.
\end{acknowledgements}

\bibliography{biblio}

\begin{thebibliography}{111}%
\makeatletter
\providecommand \@ifxundefined [1]{%
 \@ifx{#1\undefined}
}%
\providecommand \@ifnum [1]{%
 \ifnum #1\expandafter \@firstoftwo
 \else \expandafter \@secondoftwo
 \fi
}%
\providecommand \@ifx [1]{%
 \ifx #1\expandafter \@firstoftwo
 \else \expandafter \@secondoftwo
 \fi
}%
\providecommand \natexlab [1]{#1}%
\providecommand \enquote  [1]{``#1''}%
\providecommand \bibnamefont  [1]{#1}%
\providecommand \bibfnamefont [1]{#1}%
\providecommand \citenamefont [1]{#1}%
\providecommand \href@noop [0]{\@secondoftwo}%
\providecommand \href [0]{\begingroup \@sanitize@url \@href}%
\providecommand \@href[1]{\@@startlink{#1}\@@href}%
\providecommand \@@href[1]{\endgroup#1\@@endlink}%
\providecommand \@sanitize@url [0]{\catcode `\\12\catcode `\$12\catcode
  `\&12\catcode `\#12\catcode `\^12\catcode `\_12\catcode `\%12\relax}%
\providecommand \@@startlink[1]{}%
\providecommand \@@endlink[0]{}%
\providecommand \url  [0]{\begingroup\@sanitize@url \@url }%
\providecommand \@url [1]{\endgroup\@href {#1}{\urlprefix }}%
\providecommand \urlprefix  [0]{URL }%
\providecommand \Eprint [0]{\href }%
\providecommand \doibase [0]{http://dx.doi.org/}%
\providecommand \selectlanguage [0]{\@gobble}%
\providecommand \bibinfo  [0]{\@secondoftwo}%
\providecommand \bibfield  [0]{\@secondoftwo}%
\providecommand \translation [1]{[#1]}%
\providecommand \BibitemOpen [0]{}%
\providecommand \bibitemStop [0]{}%
\providecommand \bibitemNoStop [0]{.\EOS\space}%
\providecommand \EOS [0]{\spacefactor3000\relax}%
\providecommand \BibitemShut  [1]{\csname bibitem#1\endcsname}%
\let\auto@bib@innerbib\@empty
\bibitem [{\citenamefont {Mori}\ \emph {et~al.}(2018)\citenamefont {Mori},
  \citenamefont {Ikeda}, \citenamefont {Kaminishi},\ and\ \citenamefont
  {Ueda}}]{Mori_review}%
  \BibitemOpen
  \bibfield  {author} {\bibinfo {author} {\bibfnamefont {Takashi}\ \bibnamefont
  {Mori}}, \bibinfo {author} {\bibfnamefont {Tatsuhiko~N}\ \bibnamefont
  {Ikeda}}, \bibinfo {author} {\bibfnamefont {Eriko}\ \bibnamefont
  {Kaminishi}}, \ and\ \bibinfo {author} {\bibfnamefont {Masahito}\
  \bibnamefont {Ueda}},\ }\bibfield  {title} {\enquote {\bibinfo {title}
  {Thermalization and prethermalization in isolated quantum systems: a
  theoretical overview},}\ }\href {\doibase 10.1088/1361-6455/aabcdf} {\
  \textbf {\bibinfo {volume} {51}},\ \bibinfo {pages} {112001} (\bibinfo {year}
  {2018})}\BibitemShut {NoStop}%
\bibitem [{\citenamefont {Heyl}(2018)}]{Heyl_review}%
  \BibitemOpen
  \bibfield  {author} {\bibinfo {author} {\bibfnamefont {Markus}\ \bibnamefont
  {Heyl}},\ }\bibfield  {title} {\enquote {\bibinfo {title} {Dynamical quantum
  phase transitions: a review},}\ }\href {\doibase 10.1088/1361-6633/aaaf9a}
  {\bibfield  {journal} {\bibinfo  {journal} {Reports on Progress in Physics}\
  }\textbf {\bibinfo {volume} {81}},\ \bibinfo {pages} {054001} (\bibinfo
  {year} {2018})}\BibitemShut {NoStop}%
\bibitem [{\citenamefont {Cardy}(1996)}]{Cardy_book}%
  \BibitemOpen
  \bibfield  {author} {\bibinfo {author} {\bibfnamefont {J.}~\bibnamefont
  {Cardy}},\ }\href {https://books.google.de/books?id=Wt804S9FjyAC} {\emph
  {\bibinfo {title} {Scaling and Renormalization in Statistical Physics}}},\
  Cambridge Lecture Notes in Physics\ (\bibinfo  {publisher} {Cambridge
  University Press},\ \bibinfo {year} {1996})\BibitemShut {NoStop}%
\bibitem [{\citenamefont {Moeckel}\ and\ \citenamefont
  {Kehrein}(2008)}]{Moeckel2008}%
  \BibitemOpen
  \bibfield  {author} {\bibinfo {author} {\bibfnamefont {Michael}\ \bibnamefont
  {Moeckel}}\ and\ \bibinfo {author} {\bibfnamefont {Stefan}\ \bibnamefont
  {Kehrein}},\ }\bibfield  {title} {\enquote {\bibinfo {title} {Interaction
  quench in the hubbard model},}\ }\href {\doibase
  10.1103/PhysRevLett.100.175702} {\bibfield  {journal} {\bibinfo  {journal}
  {Phys. Rev. Lett.}\ }\textbf {\bibinfo {volume} {100}},\ \bibinfo {pages}
  {175702} (\bibinfo {year} {2008})}\BibitemShut {NoStop}%
\bibitem [{\citenamefont {Eckstein}\ and\ \citenamefont
  {Kollar}(2008)}]{Eckstein2008}%
  \BibitemOpen
  \bibfield  {author} {\bibinfo {author} {\bibfnamefont {Martin}\ \bibnamefont
  {Eckstein}}\ and\ \bibinfo {author} {\bibfnamefont {Marcus}\ \bibnamefont
  {Kollar}},\ }\bibfield  {title} {\enquote {\bibinfo {title} {Nonthermal
  steady states after an interaction quench in the falicov-kimball model},}\
  }\href {\doibase 10.1103/PhysRevLett.100.120404} {\bibfield  {journal}
  {\bibinfo  {journal} {Phys. Rev. Lett.}\ }\textbf {\bibinfo {volume} {100}},\
  \bibinfo {pages} {120404} (\bibinfo {year} {2008})}\BibitemShut {NoStop}%
\bibitem [{\citenamefont {Sciolla}\ and\ \citenamefont
  {Biroli}(2010)}]{Sciolla2010}%
  \BibitemOpen
  \bibfield  {author} {\bibinfo {author} {\bibfnamefont {Bruno}\ \bibnamefont
  {Sciolla}}\ and\ \bibinfo {author} {\bibfnamefont {Giulio}\ \bibnamefont
  {Biroli}},\ }\bibfield  {title} {\enquote {\bibinfo {title} {Quantum quenches
  and off-equilibrium dynamical transition in the infinite-dimensional
  bose-hubbard model},}\ }\href {\doibase 10.1103/PhysRevLett.105.220401}
  {\bibfield  {journal} {\bibinfo  {journal} {Phys. Rev. Lett.}\ }\textbf
  {\bibinfo {volume} {105}},\ \bibinfo {pages} {220401} (\bibinfo {year}
  {2010})}\BibitemShut {NoStop}%
\bibitem [{\citenamefont {Sciolla}\ and\ \citenamefont
  {Biroli}(2011)}]{Sciolla2011}%
  \BibitemOpen
  \bibfield  {author} {\bibinfo {author} {\bibfnamefont {Bruno}\ \bibnamefont
  {Sciolla}}\ and\ \bibinfo {author} {\bibfnamefont {Giulio}\ \bibnamefont
  {Biroli}},\ }\bibfield  {title} {\enquote {\bibinfo {title} {Dynamical
  transitions and quantum quenches in mean-field models},}\ }\href {\doibase
  10.1088/1742-5468/2011/11/P11003} {\bibfield  {journal} {\bibinfo  {journal}
  {Journal of Statistical Mechanics: Theory and Experiment}\ }\textbf {\bibinfo
  {volume} {2011}},\ \bibinfo {pages} {P11003} (\bibinfo {year}
  {2011})}\BibitemShut {NoStop}%
\bibitem [{\citenamefont {Žunkovič}\ \emph {et~al.}(2016)\citenamefont
  {Žunkovič}, \citenamefont {Silva},\ and\ \citenamefont
  {Fabrizio}}]{Zunkovic2016}%
  \BibitemOpen
  \bibfield  {author} {\bibinfo {author} {\bibfnamefont {Bojan}\ \bibnamefont
  {Žunkovič}}, \bibinfo {author} {\bibfnamefont {Alessandro}\ \bibnamefont
  {Silva}}, \ and\ \bibinfo {author} {\bibfnamefont {Michele}\ \bibnamefont
  {Fabrizio}},\ }\bibfield  {title} {\enquote {\bibinfo {title} {Dynamical
  phase transitions and loschmidt echo in the infinite-range xy model},}\
  }\href {\doibase 10.1098/rsta.2015.0160} {\bibfield  {journal} {\bibinfo
  {journal} {Philosophical Transactions of the Royal Society A: Mathematical,
  Physical and Engineering Sciences}\ }\textbf {\bibinfo {volume} {374}},\
  \bibinfo {pages} {20150160} (\bibinfo {year} {2016})}\BibitemShut {NoStop}%
\bibitem [{\citenamefont {Homrighausen}\ \emph {et~al.}(2017)\citenamefont
  {Homrighausen}, \citenamefont {Abeling}, \citenamefont {Zauner-Stauber},\
  and\ \citenamefont {Halimeh}}]{Homrighausen2017anomalous}%
  \BibitemOpen
  \bibfield  {author} {\bibinfo {author} {\bibfnamefont {Ingo}\ \bibnamefont
  {Homrighausen}}, \bibinfo {author} {\bibfnamefont {Nils~O.}\ \bibnamefont
  {Abeling}}, \bibinfo {author} {\bibfnamefont {Valentin}\ \bibnamefont
  {Zauner-Stauber}}, \ and\ \bibinfo {author} {\bibfnamefont {Jad~C.}\
  \bibnamefont {Halimeh}},\ }\bibfield  {title} {\enquote {\bibinfo {title}
  {Anomalous dynamical phase in quantum spin chains with long-range
  interactions},}\ }\href {\doibase 10.1103/PhysRevB.96.104436} {\bibfield
  {journal} {\bibinfo  {journal} {Phys. Rev. B}\ }\textbf {\bibinfo {volume}
  {96}},\ \bibinfo {pages} {104436} (\bibinfo {year} {2017})}\BibitemShut
  {NoStop}%
\bibitem [{\citenamefont {Lang}\ \emph
  {et~al.}(2018{\natexlab{a}})\citenamefont {Lang}, \citenamefont {Frank},\
  and\ \citenamefont {Halimeh}}]{Lang2018concurrence}%
  \BibitemOpen
  \bibfield  {author} {\bibinfo {author} {\bibfnamefont {Johannes}\
  \bibnamefont {Lang}}, \bibinfo {author} {\bibfnamefont {Bernhard}\
  \bibnamefont {Frank}}, \ and\ \bibinfo {author} {\bibfnamefont {Jad~C.}\
  \bibnamefont {Halimeh}},\ }\bibfield  {title} {\enquote {\bibinfo {title}
  {Concurrence of dynamical phase transitions at finite temperature in the
  fully connected transverse-field ising model},}\ }\href {\doibase
  10.1103/PhysRevB.97.174401} {\bibfield  {journal} {\bibinfo  {journal} {Phys.
  Rev. B}\ }\textbf {\bibinfo {volume} {97}},\ \bibinfo {pages} {174401}
  (\bibinfo {year} {2018}{\natexlab{a}})}\BibitemShut {NoStop}%
\bibitem [{\citenamefont {Lang}\ \emph
  {et~al.}(2018{\natexlab{b}})\citenamefont {Lang}, \citenamefont {Frank},\
  and\ \citenamefont {Halimeh}}]{Lang2018geometric}%
  \BibitemOpen
  \bibfield  {author} {\bibinfo {author} {\bibfnamefont {Johannes}\
  \bibnamefont {Lang}}, \bibinfo {author} {\bibfnamefont {Bernhard}\
  \bibnamefont {Frank}}, \ and\ \bibinfo {author} {\bibfnamefont {Jad~C.}\
  \bibnamefont {Halimeh}},\ }\bibfield  {title} {\enquote {\bibinfo {title}
  {Dynamical quantum phase transitions: A geometric picture},}\ }\href
  {\doibase 10.1103/PhysRevLett.121.130603} {\bibfield  {journal} {\bibinfo
  {journal} {Phys. Rev. Lett.}\ }\textbf {\bibinfo {volume} {121}},\ \bibinfo
  {pages} {130603} (\bibinfo {year} {2018}{\natexlab{b}})}\BibitemShut
  {NoStop}%
\bibitem [{\citenamefont {Corps}\ and\ \citenamefont
  {Rela\~no}(2022)}]{Corps2022dynamical}%
  \BibitemOpen
  \bibfield  {author} {\bibinfo {author} {\bibfnamefont {\'Angel~L.}\
  \bibnamefont {Corps}}\ and\ \bibinfo {author} {\bibfnamefont {Armando}\
  \bibnamefont {Rela\~no}},\ }\bibfield  {title} {\enquote {\bibinfo {title}
  {Dynamical and excited-state quantum phase transitions in collective
  systems},}\ }\href {\doibase 10.1103/PhysRevB.106.024311} {\bibfield
  {journal} {\bibinfo  {journal} {Phys. Rev. B}\ }\textbf {\bibinfo {volume}
  {106}},\ \bibinfo {pages} {024311} (\bibinfo {year} {2022})}\BibitemShut
  {NoStop}%
\bibitem [{\citenamefont {Corps}\ and\ \citenamefont
  {Rela\~no}(2023)}]{Corps2023theory}%
  \BibitemOpen
  \bibfield  {author} {\bibinfo {author} {\bibfnamefont {\'Angel~L.}\
  \bibnamefont {Corps}}\ and\ \bibinfo {author} {\bibfnamefont {Armando}\
  \bibnamefont {Rela\~no}},\ }\bibfield  {title} {\enquote {\bibinfo {title}
  {Theory of dynamical phase transitions in quantum systems with
  symmetry-breaking eigenstates},}\ }\href {\doibase
  10.1103/PhysRevLett.130.100402} {\bibfield  {journal} {\bibinfo  {journal}
  {Phys. Rev. Lett.}\ }\textbf {\bibinfo {volume} {130}},\ \bibinfo {pages}
  {100402} (\bibinfo {year} {2023})}\BibitemShut {NoStop}%
\bibitem [{\citenamefont {Corps}\ \emph {et~al.}(2023)\citenamefont {Corps},
  \citenamefont {P\'erez-Fern\'andez},\ and\ \citenamefont
  {Rela\~no}}]{Corps2023relaxation}%
  \BibitemOpen
  \bibfield  {author} {\bibinfo {author} {\bibfnamefont {\'Angel~L.}\
  \bibnamefont {Corps}}, \bibinfo {author} {\bibfnamefont {Pedro}\ \bibnamefont
  {P\'erez-Fern\'andez}}, \ and\ \bibinfo {author} {\bibfnamefont {Armando}\
  \bibnamefont {Rela\~no}},\ }\bibfield  {title} {\enquote {\bibinfo {title}
  {Relaxation time as a control parameter for exploring dynamical phase
  diagrams},}\ }\href {\doibase 10.1103/PhysRevB.108.174305} {\bibfield
  {journal} {\bibinfo  {journal} {Phys. Rev. B}\ }\textbf {\bibinfo {volume}
  {108}},\ \bibinfo {pages} {174305} (\bibinfo {year} {2023})}\BibitemShut
  {NoStop}%
\bibitem [{\citenamefont {Eckstein}\ \emph {et~al.}(2009)\citenamefont
  {Eckstein}, \citenamefont {Kollar},\ and\ \citenamefont
  {Werner}}]{Eckstein2009}%
  \BibitemOpen
  \bibfield  {author} {\bibinfo {author} {\bibfnamefont {Martin}\ \bibnamefont
  {Eckstein}}, \bibinfo {author} {\bibfnamefont {Marcus}\ \bibnamefont
  {Kollar}}, \ and\ \bibinfo {author} {\bibfnamefont {Philipp}\ \bibnamefont
  {Werner}},\ }\bibfield  {title} {\enquote {\bibinfo {title} {Thermalization
  after an interaction quench in the hubbard model},}\ }\href {\doibase
  10.1103/PhysRevLett.103.056403} {\bibfield  {journal} {\bibinfo  {journal}
  {Phys. Rev. Lett.}\ }\textbf {\bibinfo {volume} {103}},\ \bibinfo {pages}
  {056403} (\bibinfo {year} {2009})}\BibitemShut {NoStop}%
\bibitem [{\citenamefont {Moeckel}\ and\ \citenamefont
  {Kehrein}(2010)}]{Moeckel2010}%
  \BibitemOpen
  \bibfield  {author} {\bibinfo {author} {\bibfnamefont {Michael}\ \bibnamefont
  {Moeckel}}\ and\ \bibinfo {author} {\bibfnamefont {Stefan}\ \bibnamefont
  {Kehrein}},\ }\bibfield  {title} {\enquote {\bibinfo {title} {Crossover from
  adiabatic to sudden interaction quenches in the hubbard model:
  prethermalization and non-equilibrium dynamics},}\ }\href {\doibase
  10.1088/1367-2630/12/5/055016} {\bibfield  {journal} {\bibinfo  {journal}
  {New Journal of Physics}\ }\textbf {\bibinfo {volume} {12}},\ \bibinfo
  {pages} {055016} (\bibinfo {year} {2010})}\BibitemShut {NoStop}%
\bibitem [{\citenamefont {Tsuji}\ \emph {et~al.}(2013)\citenamefont {Tsuji},
  \citenamefont {Eckstein},\ and\ \citenamefont {Werner}}]{Tsuji2013}%
  \BibitemOpen
  \bibfield  {author} {\bibinfo {author} {\bibfnamefont {Naoto}\ \bibnamefont
  {Tsuji}}, \bibinfo {author} {\bibfnamefont {Martin}\ \bibnamefont
  {Eckstein}}, \ and\ \bibinfo {author} {\bibfnamefont {Philipp}\ \bibnamefont
  {Werner}},\ }\bibfield  {title} {\enquote {\bibinfo {title} {Nonthermal
  antiferromagnetic order and nonequilibrium criticality in the hubbard
  model},}\ }\href {\doibase 10.1103/PhysRevLett.110.136404} {\bibfield
  {journal} {\bibinfo  {journal} {Phys. Rev. Lett.}\ }\textbf {\bibinfo
  {volume} {110}},\ \bibinfo {pages} {136404} (\bibinfo {year}
  {2013})}\BibitemShut {NoStop}%
\bibitem [{\citenamefont {Chandran}\ \emph {et~al.}(2013)\citenamefont
  {Chandran}, \citenamefont {Nanduri}, \citenamefont {Gubser},\ and\
  \citenamefont {Sondhi}}]{Chandan2013}%
  \BibitemOpen
  \bibfield  {author} {\bibinfo {author} {\bibfnamefont {Anushya}\ \bibnamefont
  {Chandran}}, \bibinfo {author} {\bibfnamefont {Arun}\ \bibnamefont
  {Nanduri}}, \bibinfo {author} {\bibfnamefont {S.~S.}\ \bibnamefont {Gubser}},
  \ and\ \bibinfo {author} {\bibfnamefont {S.~L.}\ \bibnamefont {Sondhi}},\
  }\bibfield  {title} {\enquote {\bibinfo {title} {Equilibration and coarsening
  in the quantum $o(n)$ model at infinite $n$},}\ }\href {\doibase
  10.1103/PhysRevB.88.024306} {\bibfield  {journal} {\bibinfo  {journal} {Phys.
  Rev. B}\ }\textbf {\bibinfo {volume} {88}},\ \bibinfo {pages} {024306}
  (\bibinfo {year} {2013})}\BibitemShut {NoStop}%
\bibitem [{\citenamefont {Maraga}\ \emph {et~al.}(2015)\citenamefont {Maraga},
  \citenamefont {Chiocchetta}, \citenamefont {Mitra},\ and\ \citenamefont
  {Gambassi}}]{Maraga2015}%
  \BibitemOpen
  \bibfield  {author} {\bibinfo {author} {\bibfnamefont {Anna}\ \bibnamefont
  {Maraga}}, \bibinfo {author} {\bibfnamefont {Alessio}\ \bibnamefont
  {Chiocchetta}}, \bibinfo {author} {\bibfnamefont {Aditi}\ \bibnamefont
  {Mitra}}, \ and\ \bibinfo {author} {\bibfnamefont {Andrea}\ \bibnamefont
  {Gambassi}},\ }\bibfield  {title} {\enquote {\bibinfo {title} {Aging and
  coarsening in isolated quantum systems after a quench: Exact results for the
  quantum $\text{O}(n)$ model with $n$ $\ensuremath{\rightarrow}$
  $\ensuremath{\infty}$},}\ }\href {\doibase 10.1103/PhysRevE.92.042151}
  {\bibfield  {journal} {\bibinfo  {journal} {Phys. Rev. E}\ }\textbf {\bibinfo
  {volume} {92}},\ \bibinfo {pages} {042151} (\bibinfo {year}
  {2015})}\BibitemShut {NoStop}%
\bibitem [{\citenamefont {Smacchia}\ \emph {et~al.}(2015)\citenamefont
  {Smacchia}, \citenamefont {Knap}, \citenamefont {Demler},\ and\ \citenamefont
  {Silva}}]{Smacchia2015}%
  \BibitemOpen
  \bibfield  {author} {\bibinfo {author} {\bibfnamefont {Pietro}\ \bibnamefont
  {Smacchia}}, \bibinfo {author} {\bibfnamefont {Michael}\ \bibnamefont
  {Knap}}, \bibinfo {author} {\bibfnamefont {Eugene}\ \bibnamefont {Demler}}, \
  and\ \bibinfo {author} {\bibfnamefont {Alessandro}\ \bibnamefont {Silva}},\
  }\bibfield  {title} {\enquote {\bibinfo {title} {Exploring dynamical phase
  transitions and prethermalization with quantum noise of excitations},}\
  }\href {\doibase 10.1103/PhysRevB.91.205136} {\bibfield  {journal} {\bibinfo
  {journal} {Phys. Rev. B}\ }\textbf {\bibinfo {volume} {91}},\ \bibinfo
  {pages} {205136} (\bibinfo {year} {2015})}\BibitemShut {NoStop}%
\bibitem [{\citenamefont {Chiocchetta}\ \emph {et~al.}(2017)\citenamefont
  {Chiocchetta}, \citenamefont {Gambassi}, \citenamefont {Diehl},\ and\
  \citenamefont {Marino}}]{Chiocchetta2017}%
  \BibitemOpen
  \bibfield  {author} {\bibinfo {author} {\bibfnamefont {Alessio}\ \bibnamefont
  {Chiocchetta}}, \bibinfo {author} {\bibfnamefont {Andrea}\ \bibnamefont
  {Gambassi}}, \bibinfo {author} {\bibfnamefont {Sebastian}\ \bibnamefont
  {Diehl}}, \ and\ \bibinfo {author} {\bibfnamefont {Jamir}\ \bibnamefont
  {Marino}},\ }\bibfield  {title} {\enquote {\bibinfo {title} {Dynamical
  crossovers in prethermal critical states},}\ }\href {\doibase
  10.1103/PhysRevLett.118.135701} {\bibfield  {journal} {\bibinfo  {journal}
  {Phys. Rev. Lett.}\ }\textbf {\bibinfo {volume} {118}},\ \bibinfo {pages}
  {135701} (\bibinfo {year} {2017})}\BibitemShut {NoStop}%
\bibitem [{\citenamefont {Halimeh}\ and\ \citenamefont
  {Maghrebi}(2021)}]{Halimeh2021}%
  \BibitemOpen
  \bibfield  {author} {\bibinfo {author} {\bibfnamefont {Jad~C.}\ \bibnamefont
  {Halimeh}}\ and\ \bibinfo {author} {\bibfnamefont {Mohammad~F.}\ \bibnamefont
  {Maghrebi}},\ }\bibfield  {title} {\enquote {\bibinfo {title} {Quantum aging
  and dynamical universality in the long-range
  $o(n\ensuremath{\rightarrow}\ensuremath{\infty})$ model},}\ }\href {\doibase
  10.1103/PhysRevE.103.052142} {\bibfield  {journal} {\bibinfo  {journal}
  {Phys. Rev. E}\ }\textbf {\bibinfo {volume} {103}},\ \bibinfo {pages}
  {052142} (\bibinfo {year} {2021})}\BibitemShut {NoStop}%
\bibitem [{\citenamefont {Halimeh}\ \emph {et~al.}(2017)\citenamefont
  {Halimeh}, \citenamefont {Zauner-Stauber}, \citenamefont {McCulloch},
  \citenamefont {de~Vega}, \citenamefont {Schollw\"ock},\ and\ \citenamefont
  {Kastner}}]{Halimeh2017}%
  \BibitemOpen
  \bibfield  {author} {\bibinfo {author} {\bibfnamefont {Jad~C.}\ \bibnamefont
  {Halimeh}}, \bibinfo {author} {\bibfnamefont {Valentin}\ \bibnamefont
  {Zauner-Stauber}}, \bibinfo {author} {\bibfnamefont {Ian~P.}\ \bibnamefont
  {McCulloch}}, \bibinfo {author} {\bibfnamefont {In\'es}\ \bibnamefont
  {de~Vega}}, \bibinfo {author} {\bibfnamefont {Ulrich}\ \bibnamefont
  {Schollw\"ock}}, \ and\ \bibinfo {author} {\bibfnamefont {Michael}\
  \bibnamefont {Kastner}},\ }\bibfield  {title} {\enquote {\bibinfo {title}
  {Prethermalization and persistent order in the absence of a thermal phase
  transition},}\ }\href {\doibase 10.1103/PhysRevB.95.024302} {\bibfield
  {journal} {\bibinfo  {journal} {Phys. Rev. B}\ }\textbf {\bibinfo {volume}
  {95}},\ \bibinfo {pages} {024302} (\bibinfo {year} {2017})}\BibitemShut
  {NoStop}%
\bibitem [{\citenamefont {\ifmmode \check{Z}\else
  \v{Z}\fi{}unkovi\ifmmode~\check{c}\else \v{c}\fi{}}\ \emph
  {et~al.}(2018)\citenamefont {\ifmmode \check{Z}\else
  \v{Z}\fi{}unkovi\ifmmode~\check{c}\else \v{c}\fi{}}, \citenamefont {Heyl},
  \citenamefont {Knap},\ and\ \citenamefont {Silva}}]{Zunkovic2018dynamical}%
  \BibitemOpen
  \bibfield  {author} {\bibinfo {author} {\bibfnamefont {Bojan}\ \bibnamefont
  {\ifmmode \check{Z}\else \v{Z}\fi{}unkovi\ifmmode~\check{c}\else
  \v{c}\fi{}}}, \bibinfo {author} {\bibfnamefont {Markus}\ \bibnamefont
  {Heyl}}, \bibinfo {author} {\bibfnamefont {Michael}\ \bibnamefont {Knap}}, \
  and\ \bibinfo {author} {\bibfnamefont {Alessandro}\ \bibnamefont {Silva}},\
  }\bibfield  {title} {\enquote {\bibinfo {title} {Dynamical quantum phase
  transitions in spin chains with long-range interactions: Merging different
  concepts of nonequilibrium criticality},}\ }\href {\doibase
  10.1103/PhysRevLett.120.130601} {\bibfield  {journal} {\bibinfo  {journal}
  {Phys. Rev. Lett.}\ }\textbf {\bibinfo {volume} {120}},\ \bibinfo {pages}
  {130601} (\bibinfo {year} {2018})}\BibitemShut {NoStop}%
\bibitem [{\citenamefont {Gambassi}\ and\ \citenamefont
  {Calabrese}(2011)}]{Gambassi2011}%
  \BibitemOpen
  \bibfield  {author} {\bibinfo {author} {\bibfnamefont {A.}~\bibnamefont
  {Gambassi}}\ and\ \bibinfo {author} {\bibfnamefont {P.}~\bibnamefont
  {Calabrese}},\ }\bibfield  {title} {\enquote {\bibinfo {title} {Quantum
  quenches as classical critical films},}\ }\href {\doibase
  10.1209/0295-5075/95/66007} {\bibfield  {journal} {\bibinfo  {journal}
  {Europhysics Letters}\ }\textbf {\bibinfo {volume} {95}},\ \bibinfo {pages}
  {66007} (\bibinfo {year} {2011})}\BibitemShut {NoStop}%
\bibitem [{\citenamefont {Langen}\ \emph {et~al.}(2016)\citenamefont {Langen},
  \citenamefont {Gasenzer},\ and\ \citenamefont {Schmiedmayer}}]{Langen2016}%
  \BibitemOpen
  \bibfield  {author} {\bibinfo {author} {\bibfnamefont {Tim}\ \bibnamefont
  {Langen}}, \bibinfo {author} {\bibfnamefont {Thomas}\ \bibnamefont
  {Gasenzer}}, \ and\ \bibinfo {author} {\bibfnamefont {Jörg}\ \bibnamefont
  {Schmiedmayer}},\ }\bibfield  {title} {\enquote {\bibinfo {title}
  {Prethermalization and universal dynamics in near-integrable quantum
  systems},}\ }\href {\doibase 10.1088/1742-5468/2016/06/064009} {\bibfield
  {journal} {\bibinfo  {journal} {Journal of Statistical Mechanics: Theory and
  Experiment}\ }\textbf {\bibinfo {volume} {2016}},\ \bibinfo {pages} {064009}
  (\bibinfo {year} {2016})}\BibitemShut {NoStop}%
\bibitem [{\citenamefont {Marcuzzi}\ \emph {et~al.}(2016)\citenamefont
  {Marcuzzi}, \citenamefont {Marino}, \citenamefont {Gambassi},\ and\
  \citenamefont {Silva}}]{Marcuzzi2016}%
  \BibitemOpen
  \bibfield  {author} {\bibinfo {author} {\bibfnamefont {M.}~\bibnamefont
  {Marcuzzi}}, \bibinfo {author} {\bibfnamefont {J.}~\bibnamefont {Marino}},
  \bibinfo {author} {\bibfnamefont {A.}~\bibnamefont {Gambassi}}, \ and\
  \bibinfo {author} {\bibfnamefont {A.}~\bibnamefont {Silva}},\ }\bibfield
  {title} {\enquote {\bibinfo {title} {Prethermalization from a low-density
  holstein-primakoff expansion},}\ }\href {\doibase 10.1103/PhysRevB.94.214304}
  {\bibfield  {journal} {\bibinfo  {journal} {Phys. Rev. B}\ }\textbf {\bibinfo
  {volume} {94}},\ \bibinfo {pages} {214304} (\bibinfo {year}
  {2016})}\BibitemShut {NoStop}%
\bibitem [{\citenamefont {Heyl}\ \emph {et~al.}(2013)\citenamefont {Heyl},
  \citenamefont {Polkovnikov},\ and\ \citenamefont {Kehrein}}]{Heyl2013}%
  \BibitemOpen
  \bibfield  {author} {\bibinfo {author} {\bibfnamefont {M.}~\bibnamefont
  {Heyl}}, \bibinfo {author} {\bibfnamefont {A.}~\bibnamefont {Polkovnikov}}, \
  and\ \bibinfo {author} {\bibfnamefont {S.}~\bibnamefont {Kehrein}},\
  }\bibfield  {title} {\enquote {\bibinfo {title} {Dynamical quantum phase
  transitions in the transverse-field ising model},}\ }\href {\doibase
  10.1103/PhysRevLett.110.135704} {\bibfield  {journal} {\bibinfo  {journal}
  {Phys. Rev. Lett.}\ }\textbf {\bibinfo {volume} {110}},\ \bibinfo {pages}
  {135704} (\bibinfo {year} {2013})}\BibitemShut {NoStop}%
\bibitem [{\citenamefont {Heyl}(2014)}]{Heyl2014}%
  \BibitemOpen
  \bibfield  {author} {\bibinfo {author} {\bibfnamefont {M.}~\bibnamefont
  {Heyl}},\ }\bibfield  {title} {\enquote {\bibinfo {title} {Dynamical quantum
  phase transitions in systems with broken-symmetry phases},}\ }\href {\doibase
  10.1103/PhysRevLett.113.205701} {\bibfield  {journal} {\bibinfo  {journal}
  {Phys. Rev. Lett.}\ }\textbf {\bibinfo {volume} {113}},\ \bibinfo {pages}
  {205701} (\bibinfo {year} {2014})}\BibitemShut {NoStop}%
\bibitem [{\citenamefont {Heyl}(2015)}]{Heyl2015}%
  \BibitemOpen
  \bibfield  {author} {\bibinfo {author} {\bibfnamefont {Markus}\ \bibnamefont
  {Heyl}},\ }\bibfield  {title} {\enquote {\bibinfo {title} {Scaling and
  universality at dynamical quantum phase transitions},}\ }\href {\doibase
  10.1103/PhysRevLett.115.140602} {\bibfield  {journal} {\bibinfo  {journal}
  {Phys. Rev. Lett.}\ }\textbf {\bibinfo {volume} {115}},\ \bibinfo {pages}
  {140602} (\bibinfo {year} {2015})}\BibitemShut {NoStop}%
\bibitem [{\citenamefont {Karrasch}\ and\ \citenamefont
  {Schuricht}(2013)}]{Karrasch2013dynamical}%
  \BibitemOpen
  \bibfield  {author} {\bibinfo {author} {\bibfnamefont {C.}~\bibnamefont
  {Karrasch}}\ and\ \bibinfo {author} {\bibfnamefont {D.}~\bibnamefont
  {Schuricht}},\ }\bibfield  {title} {\enquote {\bibinfo {title} {Dynamical
  phase transitions after quenches in nonintegrable models},}\ }\href {\doibase
  10.1103/PhysRevB.87.195104} {\bibfield  {journal} {\bibinfo  {journal} {Phys.
  Rev. B}\ }\textbf {\bibinfo {volume} {87}},\ \bibinfo {pages} {195104}
  (\bibinfo {year} {2013})}\BibitemShut {NoStop}%
\bibitem [{\citenamefont {Vajna}\ and\ \citenamefont
  {D\'ora}(2014)}]{Vajna2014disentangling}%
  \BibitemOpen
  \bibfield  {author} {\bibinfo {author} {\bibfnamefont {Szabolcs}\
  \bibnamefont {Vajna}}\ and\ \bibinfo {author} {\bibfnamefont {Bal\'azs}\
  \bibnamefont {D\'ora}},\ }\bibfield  {title} {\enquote {\bibinfo {title}
  {Disentangling dynamical phase transitions from equilibrium phase
  transitions},}\ }\href {\doibase 10.1103/PhysRevB.89.161105} {\bibfield
  {journal} {\bibinfo  {journal} {Phys. Rev. B}\ }\textbf {\bibinfo {volume}
  {89}},\ \bibinfo {pages} {161105} (\bibinfo {year} {2014})}\BibitemShut
  {NoStop}%
\bibitem [{\citenamefont {Andraschko}\ and\ \citenamefont
  {Sirker}(2014)}]{Andraschko2014dynamical}%
  \BibitemOpen
  \bibfield  {author} {\bibinfo {author} {\bibfnamefont {F.}~\bibnamefont
  {Andraschko}}\ and\ \bibinfo {author} {\bibfnamefont {J.}~\bibnamefont
  {Sirker}},\ }\bibfield  {title} {\enquote {\bibinfo {title} {Dynamical
  quantum phase transitions and the loschmidt echo: A transfer matrix
  approach},}\ }\href {\doibase 10.1103/PhysRevB.89.125120} {\bibfield
  {journal} {\bibinfo  {journal} {Phys. Rev. B}\ }\textbf {\bibinfo {volume}
  {89}},\ \bibinfo {pages} {125120} (\bibinfo {year} {2014})}\BibitemShut
  {NoStop}%
\bibitem [{\citenamefont {Halimeh}\ and\ \citenamefont
  {Zauner-Stauber}(2017)}]{Halimeh2017dynamical}%
  \BibitemOpen
  \bibfield  {author} {\bibinfo {author} {\bibfnamefont {Jad~C.}\ \bibnamefont
  {Halimeh}}\ and\ \bibinfo {author} {\bibfnamefont {Valentin}\ \bibnamefont
  {Zauner-Stauber}},\ }\bibfield  {title} {\enquote {\bibinfo {title}
  {Dynamical phase diagram of quantum spin chains with long-range
  interactions},}\ }\href {\doibase 10.1103/PhysRevB.96.134427} {\bibfield
  {journal} {\bibinfo  {journal} {Phys. Rev. B}\ }\textbf {\bibinfo {volume}
  {96}},\ \bibinfo {pages} {134427} (\bibinfo {year} {2017})}\BibitemShut
  {NoStop}%
\bibitem [{\citenamefont {Zauner-Stauber}\ and\ \citenamefont
  {Halimeh}(2017)}]{Zauner-Stauber2017probing}%
  \BibitemOpen
  \bibfield  {author} {\bibinfo {author} {\bibfnamefont {Valentin}\
  \bibnamefont {Zauner-Stauber}}\ and\ \bibinfo {author} {\bibfnamefont
  {Jad~C.}\ \bibnamefont {Halimeh}},\ }\bibfield  {title} {\enquote {\bibinfo
  {title} {Probing the anomalous dynamical phase in long-range quantum spin
  chains through {F}isher-zero lines},}\ }\href {\doibase
  10.1103/PhysRevE.96.062118} {\bibfield  {journal} {\bibinfo  {journal} {Phys.
  Rev. E}\ }\textbf {\bibinfo {volume} {96}},\ \bibinfo {pages} {062118}
  (\bibinfo {year} {2017})}\BibitemShut {NoStop}%
\bibitem [{\citenamefont {Defenu}\ \emph {et~al.}(2019)\citenamefont {Defenu},
  \citenamefont {Enss},\ and\ \citenamefont {Halimeh}}]{Defenu2019dynamical}%
  \BibitemOpen
  \bibfield  {author} {\bibinfo {author} {\bibfnamefont {Nicol\`o}\
  \bibnamefont {Defenu}}, \bibinfo {author} {\bibfnamefont {Tilman}\
  \bibnamefont {Enss}}, \ and\ \bibinfo {author} {\bibfnamefont {Jad~C.}\
  \bibnamefont {Halimeh}},\ }\bibfield  {title} {\enquote {\bibinfo {title}
  {Dynamical criticality and domain-wall coupling in long-range
  hamiltonians},}\ }\href {\doibase 10.1103/PhysRevB.100.014434} {\bibfield
  {journal} {\bibinfo  {journal} {Phys. Rev. B}\ }\textbf {\bibinfo {volume}
  {100}},\ \bibinfo {pages} {014434} (\bibinfo {year} {2019})}\BibitemShut
  {NoStop}%
\bibitem [{\citenamefont {Uhrich}\ \emph {et~al.}(2020)\citenamefont {Uhrich},
  \citenamefont {Defenu}, \citenamefont {Jafari},\ and\ \citenamefont
  {Halimeh}}]{Uhrich2020out}%
  \BibitemOpen
  \bibfield  {author} {\bibinfo {author} {\bibfnamefont {Philipp}\ \bibnamefont
  {Uhrich}}, \bibinfo {author} {\bibfnamefont {Nicol\`o}\ \bibnamefont
  {Defenu}}, \bibinfo {author} {\bibfnamefont {Rouhollah}\ \bibnamefont
  {Jafari}}, \ and\ \bibinfo {author} {\bibfnamefont {Jad~C.}\ \bibnamefont
  {Halimeh}},\ }\bibfield  {title} {\enquote {\bibinfo {title}
  {Out-of-equilibrium phase diagram of long-range superconductors},}\ }\href
  {\doibase 10.1103/PhysRevB.101.245148} {\bibfield  {journal} {\bibinfo
  {journal} {Phys. Rev. B}\ }\textbf {\bibinfo {volume} {101}},\ \bibinfo
  {pages} {245148} (\bibinfo {year} {2020})}\BibitemShut {NoStop}%
\bibitem [{\citenamefont {Halimeh}\ \emph
  {et~al.}(2021{\natexlab{a}})\citenamefont {Halimeh}, \citenamefont
  {Van~Damme}, \citenamefont {Guo}, \citenamefont {Lang},\ and\ \citenamefont
  {Hauke}}]{Halimeh2021dynamical}%
  \BibitemOpen
  \bibfield  {author} {\bibinfo {author} {\bibfnamefont {Jad~C.}\ \bibnamefont
  {Halimeh}}, \bibinfo {author} {\bibfnamefont {Maarten}\ \bibnamefont
  {Van~Damme}}, \bibinfo {author} {\bibfnamefont {Lingzhen}\ \bibnamefont
  {Guo}}, \bibinfo {author} {\bibfnamefont {Johannes}\ \bibnamefont {Lang}}, \
  and\ \bibinfo {author} {\bibfnamefont {Philipp}\ \bibnamefont {Hauke}},\
  }\bibfield  {title} {\enquote {\bibinfo {title} {Dynamical phase transitions
  in quantum spin models with antiferromagnetic long-range interactions},}\
  }\href {\doibase 10.1103/PhysRevB.104.115133} {\bibfield  {journal} {\bibinfo
   {journal} {Phys. Rev. B}\ }\textbf {\bibinfo {volume} {104}},\ \bibinfo
  {pages} {115133} (\bibinfo {year} {2021}{\natexlab{a}})}\BibitemShut
  {NoStop}%
\bibitem [{\citenamefont {Mitra}\ \emph {et~al.}(2023)\citenamefont {Mitra},
  \citenamefont {Albash}, \citenamefont {Blocher}, \citenamefont {Takahashi},
  \citenamefont {Miyake}, \citenamefont {Biedermann},\ and\ \citenamefont
  {Deutsch}}]{Deutsch2023macrostates}%
  \BibitemOpen
  \bibfield  {author} {\bibinfo {author} {\bibfnamefont {Anupam}\ \bibnamefont
  {Mitra}}, \bibinfo {author} {\bibfnamefont {Tameem}\ \bibnamefont {Albash}},
  \bibinfo {author} {\bibfnamefont {Philip~Daniel}\ \bibnamefont {Blocher}},
  \bibinfo {author} {\bibfnamefont {Jun}\ \bibnamefont {Takahashi}}, \bibinfo
  {author} {\bibfnamefont {Akimasa}\ \bibnamefont {Miyake}}, \bibinfo {author}
  {\bibfnamefont {Grant~W.}\ \bibnamefont {Biedermann}}, \ and\ \bibinfo
  {author} {\bibfnamefont {Ivan~H.}\ \bibnamefont {Deutsch}},\ }\bibfield
  {title} {\enquote {\bibinfo {title} {Macrostates vs. microstates in the
  classical simulation of critical phenomena in quench dynamics of 1d ising
  models},}\ }\href@noop {} {\  (\bibinfo {year} {2023})},\ \Eprint
  {http://arxiv.org/abs/2310.08567} {arXiv:2310.08567 [quant-ph]} \BibitemShut
  {NoStop}%
\bibitem [{\citenamefont {Ángel L.~Corps}\ \emph {et~al.}(2023)\citenamefont
  {Ángel L.~Corps}, \citenamefont {Stránský},\ and\ \citenamefont
  {Cejnar}}]{Corps2023mechanism}%
  \BibitemOpen
  \bibfield  {author} {\bibinfo {author} {\bibnamefont {Ángel L.~Corps}},
  \bibinfo {author} {\bibfnamefont {Pavel}\ \bibnamefont {Stránský}}, \ and\
  \bibinfo {author} {\bibfnamefont {Pavel}\ \bibnamefont {Cejnar}},\ }\bibfield
   {title} {\enquote {\bibinfo {title} {Mechanism of dynamical phase
  transitions: The complex-time survival amplitude},}\ }\href {\doibase
  10.1103/PhysRevB.107.094307} {\bibfield  {journal} {\bibinfo  {journal}
  {Physical Review B}\ }\textbf {\bibinfo {volume} {107}},\ \bibinfo {pages}
  {094307} (\bibinfo {year} {2023})}\BibitemShut {NoStop}%
\bibitem [{\citenamefont {Vajna}\ and\ \citenamefont
  {D\'ora}(2015)}]{Vajna2015topological}%
  \BibitemOpen
  \bibfield  {author} {\bibinfo {author} {\bibfnamefont {Szabolcs}\
  \bibnamefont {Vajna}}\ and\ \bibinfo {author} {\bibfnamefont {Bal\'azs}\
  \bibnamefont {D\'ora}},\ }\bibfield  {title} {\enquote {\bibinfo {title}
  {Topological classification of dynamical phase transitions},}\ }\href
  {\doibase 10.1103/PhysRevB.91.155127} {\bibfield  {journal} {\bibinfo
  {journal} {Phys. Rev. B}\ }\textbf {\bibinfo {volume} {91}},\ \bibinfo
  {pages} {155127} (\bibinfo {year} {2015})}\BibitemShut {NoStop}%
\bibitem [{\citenamefont {Schmitt}\ and\ \citenamefont
  {Kehrein}(2015)}]{Schmitt2015dynamical}%
  \BibitemOpen
  \bibfield  {author} {\bibinfo {author} {\bibfnamefont {Markus}\ \bibnamefont
  {Schmitt}}\ and\ \bibinfo {author} {\bibfnamefont {Stefan}\ \bibnamefont
  {Kehrein}},\ }\bibfield  {title} {\enquote {\bibinfo {title} {Dynamical
  quantum phase transitions in the kitaev honeycomb model},}\ }\href {\doibase
  10.1103/PhysRevB.92.075114} {\bibfield  {journal} {\bibinfo  {journal} {Phys.
  Rev. B}\ }\textbf {\bibinfo {volume} {92}},\ \bibinfo {pages} {075114}
  (\bibinfo {year} {2015})}\BibitemShut {NoStop}%
\bibitem [{\citenamefont {Sedlmayr}\ \emph
  {et~al.}(2018{\natexlab{a}})\citenamefont {Sedlmayr}, \citenamefont {Jaeger},
  \citenamefont {Maiti},\ and\ \citenamefont {Sirker}}]{Sedlmayr2018bulk}%
  \BibitemOpen
  \bibfield  {author} {\bibinfo {author} {\bibfnamefont {N.}~\bibnamefont
  {Sedlmayr}}, \bibinfo {author} {\bibfnamefont {P.}~\bibnamefont {Jaeger}},
  \bibinfo {author} {\bibfnamefont {M.}~\bibnamefont {Maiti}}, \ and\ \bibinfo
  {author} {\bibfnamefont {J.}~\bibnamefont {Sirker}},\ }\bibfield  {title}
  {\enquote {\bibinfo {title} {Bulk-boundary correspondence for dynamical phase
  transitions in one-dimensional topological insulators and superconductors},}\
  }\href {\doibase 10.1103/PhysRevB.97.064304} {\bibfield  {journal} {\bibinfo
  {journal} {Phys. Rev. B}\ }\textbf {\bibinfo {volume} {97}},\ \bibinfo
  {pages} {064304} (\bibinfo {year} {2018}{\natexlab{a}})}\BibitemShut
  {NoStop}%
\bibitem [{\citenamefont {Hagym\'asi}\ \emph {et~al.}(2019)\citenamefont
  {Hagym\'asi}, \citenamefont {Hubig}, \citenamefont {Legeza},\ and\
  \citenamefont {Schollw\"ock}}]{Hagymasi2019dynamical}%
  \BibitemOpen
  \bibfield  {author} {\bibinfo {author} {\bibfnamefont {I.}~\bibnamefont
  {Hagym\'asi}}, \bibinfo {author} {\bibfnamefont {C.}~\bibnamefont {Hubig}},
  \bibinfo {author} {\bibfnamefont {\"O.}\ \bibnamefont {Legeza}}, \ and\
  \bibinfo {author} {\bibfnamefont {U.}~\bibnamefont {Schollw\"ock}},\
  }\bibfield  {title} {\enquote {\bibinfo {title} {Dynamical topological
  quantum phase transitions in nonintegrable models},}\ }\href {\doibase
  10.1103/PhysRevLett.122.250601} {\bibfield  {journal} {\bibinfo  {journal}
  {Phys. Rev. Lett.}\ }\textbf {\bibinfo {volume} {122}},\ \bibinfo {pages}
  {250601} (\bibinfo {year} {2019})}\BibitemShut {NoStop}%
\bibitem [{\citenamefont {Mas\l{}owski}\ and\ \citenamefont
  {Sedlmayr}(2020)}]{Maslowski2020quasiperiodic}%
  \BibitemOpen
  \bibfield  {author} {\bibinfo {author} {\bibfnamefont {T.}~\bibnamefont
  {Mas\l{}owski}}\ and\ \bibinfo {author} {\bibfnamefont {N.}~\bibnamefont
  {Sedlmayr}},\ }\bibfield  {title} {\enquote {\bibinfo {title} {Quasiperiodic
  dynamical quantum phase transitions in multiband topological insulators and
  connections with entanglement entropy and fidelity susceptibility},}\ }\href
  {\doibase 10.1103/PhysRevB.101.014301} {\bibfield  {journal} {\bibinfo
  {journal} {Phys. Rev. B}\ }\textbf {\bibinfo {volume} {101}},\ \bibinfo
  {pages} {014301} (\bibinfo {year} {2020})}\BibitemShut {NoStop}%
\bibitem [{\citenamefont {Porta}\ \emph {et~al.}(2020)\citenamefont {Porta},
  \citenamefont {Cavaliere}, \citenamefont {Sassetti},\ and\ \citenamefont
  {Traverso~Ziani}}]{Porta2020topological}%
  \BibitemOpen
  \bibfield  {author} {\bibinfo {author} {\bibfnamefont {Sergio}\ \bibnamefont
  {Porta}}, \bibinfo {author} {\bibfnamefont {Fabio}\ \bibnamefont
  {Cavaliere}}, \bibinfo {author} {\bibfnamefont {Maura}\ \bibnamefont
  {Sassetti}}, \ and\ \bibinfo {author} {\bibfnamefont {Niccol{\`o}}\
  \bibnamefont {Traverso~Ziani}},\ }\bibfield  {title} {\enquote {\bibinfo
  {title} {Topological classification of dynamical quantum phase transitions in
  the xy chain},}\ }\href {\doibase 10.1038/s41598-020-69621-8} {\bibfield
  {journal} {\bibinfo  {journal} {Scientific Reports}\ }\textbf {\bibinfo
  {volume} {10}},\ \bibinfo {pages} {12766} (\bibinfo {year}
  {2020})}\BibitemShut {NoStop}%
\bibitem [{\citenamefont {Okugawa}\ \emph {et~al.}(2021)\citenamefont
  {Okugawa}, \citenamefont {Oshiyama},\ and\ \citenamefont
  {Ohzeki}}]{Okugawa2021mirror}%
  \BibitemOpen
  \bibfield  {author} {\bibinfo {author} {\bibfnamefont {Ryo}\ \bibnamefont
  {Okugawa}}, \bibinfo {author} {\bibfnamefont {Hiroki}\ \bibnamefont
  {Oshiyama}}, \ and\ \bibinfo {author} {\bibfnamefont {Masayuki}\ \bibnamefont
  {Ohzeki}},\ }\bibfield  {title} {\enquote {\bibinfo {title}
  {Mirror-symmetry-protected dynamical quantum phase transitions in topological
  crystalline insulators},}\ }\href {\doibase 10.1103/PhysRevResearch.3.043064}
  {\bibfield  {journal} {\bibinfo  {journal} {Phys. Rev. Res.}\ }\textbf
  {\bibinfo {volume} {3}},\ \bibinfo {pages} {043064} (\bibinfo {year}
  {2021})}\BibitemShut {NoStop}%
\bibitem [{\citenamefont {Wrze\ifmmode~\acute{s}\else \'{s}\fi{}niewski}\ \emph
  {et~al.}(2022)\citenamefont {Wrze\ifmmode~\acute{s}\else \'{s}\fi{}niewski},
  \citenamefont {Weymann}, \citenamefont {Sedlmayr},\ and\ \citenamefont
  {Doma\ifmmode~\acute{n}\else \'{n}\fi{}ski}}]{Sedlmayr2022dynamical}%
  \BibitemOpen
  \bibfield  {author} {\bibinfo {author} {\bibfnamefont {K.}~\bibnamefont
  {Wrze\ifmmode~\acute{s}\else \'{s}\fi{}niewski}}, \bibinfo {author}
  {\bibfnamefont {I.}~\bibnamefont {Weymann}}, \bibinfo {author} {\bibfnamefont
  {N.}~\bibnamefont {Sedlmayr}}, \ and\ \bibinfo {author} {\bibfnamefont
  {T.}~\bibnamefont {Doma\ifmmode~\acute{n}\else \'{n}\fi{}ski}},\ }\bibfield
  {title} {\enquote {\bibinfo {title} {Dynamical quantum phase transitions in a
  mesoscopic superconducting system},}\ }\href {\doibase
  10.1103/PhysRevB.105.094514} {\bibfield  {journal} {\bibinfo  {journal}
  {Phys. Rev. B}\ }\textbf {\bibinfo {volume} {105}},\ \bibinfo {pages}
  {094514} (\bibinfo {year} {2022})}\BibitemShut {NoStop}%
\bibitem [{\citenamefont {Mas\l{}owski}\ and\ \citenamefont
  {Sedlmayr}(2023)}]{Maslowski2023dynamical}%
  \BibitemOpen
  \bibfield  {author} {\bibinfo {author} {\bibfnamefont {T.}~\bibnamefont
  {Mas\l{}owski}}\ and\ \bibinfo {author} {\bibfnamefont {N.}~\bibnamefont
  {Sedlmayr}},\ }\bibfield  {title} {\enquote {\bibinfo {title} {Dynamical
  bulk-boundary correspondence and dynamical quantum phase transitions in
  higher-order topological insulators},}\ }\href {\doibase
  10.1103/PhysRevB.108.094306} {\bibfield  {journal} {\bibinfo  {journal}
  {Phys. Rev. B}\ }\textbf {\bibinfo {volume} {108}},\ \bibinfo {pages}
  {094306} (\bibinfo {year} {2023})}\BibitemShut {NoStop}%
\bibitem [{\citenamefont {Bhattacharya}\ and\ \citenamefont
  {Dutta}(2017)}]{Bhattacharya2017emergent}%
  \BibitemOpen
  \bibfield  {author} {\bibinfo {author} {\bibfnamefont {Utso}\ \bibnamefont
  {Bhattacharya}}\ and\ \bibinfo {author} {\bibfnamefont {Amit}\ \bibnamefont
  {Dutta}},\ }\bibfield  {title} {\enquote {\bibinfo {title} {Emergent topology
  and dynamical quantum phase transitions in two-dimensional closed quantum
  systems},}\ }\href {\doibase 10.1103/PhysRevB.96.014302} {\bibfield
  {journal} {\bibinfo  {journal} {Phys. Rev. B}\ }\textbf {\bibinfo {volume}
  {96}},\ \bibinfo {pages} {014302} (\bibinfo {year} {2017})}\BibitemShut
  {NoStop}%
\bibitem [{\citenamefont {Weidinger}\ \emph {et~al.}(2017)\citenamefont
  {Weidinger}, \citenamefont {Heyl}, \citenamefont {Silva},\ and\ \citenamefont
  {Knap}}]{Weidinger2017dynamical}%
  \BibitemOpen
  \bibfield  {author} {\bibinfo {author} {\bibfnamefont {Simon~A.}\
  \bibnamefont {Weidinger}}, \bibinfo {author} {\bibfnamefont {Markus}\
  \bibnamefont {Heyl}}, \bibinfo {author} {\bibfnamefont {Alessandro}\
  \bibnamefont {Silva}}, \ and\ \bibinfo {author} {\bibfnamefont {Michael}\
  \bibnamefont {Knap}},\ }\bibfield  {title} {\enquote {\bibinfo {title}
  {Dynamical quantum phase transitions in systems with continuous symmetry
  breaking},}\ }\href {\doibase 10.1103/PhysRevB.96.134313} {\bibfield
  {journal} {\bibinfo  {journal} {Phys. Rev. B}\ }\textbf {\bibinfo {volume}
  {96}},\ \bibinfo {pages} {134313} (\bibinfo {year} {2017})}\BibitemShut
  {NoStop}%
\bibitem [{\citenamefont {Heyl}\ \emph {et~al.}(2018)\citenamefont {Heyl},
  \citenamefont {Pollmann},\ and\ \citenamefont {D\'ora}}]{Heyl2018detecting}%
  \BibitemOpen
  \bibfield  {author} {\bibinfo {author} {\bibfnamefont {Markus}\ \bibnamefont
  {Heyl}}, \bibinfo {author} {\bibfnamefont {Frank}\ \bibnamefont {Pollmann}},
  \ and\ \bibinfo {author} {\bibfnamefont {Bal\'azs}\ \bibnamefont {D\'ora}},\
  }\bibfield  {title} {\enquote {\bibinfo {title} {Detecting equilibrium and
  dynamical quantum phase transitions in ising chains via out-of-time-ordered
  correlators},}\ }\href {\doibase 10.1103/PhysRevLett.121.016801} {\bibfield
  {journal} {\bibinfo  {journal} {Phys. Rev. Lett.}\ }\textbf {\bibinfo
  {volume} {121}},\ \bibinfo {pages} {016801} (\bibinfo {year}
  {2018})}\BibitemShut {NoStop}%
\bibitem [{\citenamefont {Nicola}\ \emph {et~al.}(2019)\citenamefont {Nicola},
  \citenamefont {Doyon},\ and\ \citenamefont
  {Bhaseen}}]{DeNicola2019stochastic}%
  \BibitemOpen
  \bibfield  {author} {\bibinfo {author} {\bibfnamefont {S~De}\ \bibnamefont
  {Nicola}}, \bibinfo {author} {\bibfnamefont {B}~\bibnamefont {Doyon}}, \ and\
  \bibinfo {author} {\bibfnamefont {M~J}\ \bibnamefont {Bhaseen}},\ }\bibfield
  {title} {\enquote {\bibinfo {title} {Stochastic approach to non-equilibrium
  quantum spin systems},}\ }\href {\doibase 10.1088/1751-8121/aaf9be}
  {\bibfield  {journal} {\bibinfo  {journal} {Journal of Physics A:
  Mathematical and Theoretical}\ }\textbf {\bibinfo {volume} {52}},\ \bibinfo
  {pages} {05LT02} (\bibinfo {year} {2019})}\BibitemShut {NoStop}%
\bibitem [{\citenamefont {Hashizume}\ \emph {et~al.}(2022)\citenamefont
  {Hashizume}, \citenamefont {McCulloch},\ and\ \citenamefont
  {Halimeh}}]{Hashizume2022dynamical}%
  \BibitemOpen
  \bibfield  {author} {\bibinfo {author} {\bibfnamefont {Tomohiro}\
  \bibnamefont {Hashizume}}, \bibinfo {author} {\bibfnamefont {Ian~P.}\
  \bibnamefont {McCulloch}}, \ and\ \bibinfo {author} {\bibfnamefont {Jad~C.}\
  \bibnamefont {Halimeh}},\ }\bibfield  {title} {\enquote {\bibinfo {title}
  {Dynamical phase transitions in the two-dimensional transverse-field ising
  model},}\ }\href {\doibase 10.1103/PhysRevResearch.4.013250} {\bibfield
  {journal} {\bibinfo  {journal} {Phys. Rev. Res.}\ }\textbf {\bibinfo {volume}
  {4}},\ \bibinfo {pages} {013250} (\bibinfo {year} {2022})}\BibitemShut
  {NoStop}%
\bibitem [{\citenamefont {Hashizume}\ \emph {et~al.}(2020)\citenamefont
  {Hashizume}, \citenamefont {Halimeh},\ and\ \citenamefont
  {McCulloch}}]{Hashizume2020hybrid}%
  \BibitemOpen
  \bibfield  {author} {\bibinfo {author} {\bibfnamefont {Tomohiro}\
  \bibnamefont {Hashizume}}, \bibinfo {author} {\bibfnamefont {Jad~C.}\
  \bibnamefont {Halimeh}}, \ and\ \bibinfo {author} {\bibfnamefont {Ian~P.}\
  \bibnamefont {McCulloch}},\ }\bibfield  {title} {\enquote {\bibinfo {title}
  {Hybrid infinite time-evolving block decimation algorithm for long-range
  multidimensional quantum many-body systems},}\ }\href {\doibase
  10.1103/PhysRevB.102.035115} {\bibfield  {journal} {\bibinfo  {journal}
  {Phys. Rev. B}\ }\textbf {\bibinfo {volume} {102}},\ \bibinfo {pages}
  {035115} (\bibinfo {year} {2020})}\BibitemShut {NoStop}%
\bibitem [{\citenamefont {Kosior}\ and\ \citenamefont
  {Heyl}(2024)}]{kosior2024vortex}%
  \BibitemOpen
  \bibfield  {author} {\bibinfo {author} {\bibfnamefont {Arkadiusz}\
  \bibnamefont {Kosior}}\ and\ \bibinfo {author} {\bibfnamefont {Markus}\
  \bibnamefont {Heyl}},\ }\bibfield  {title} {\enquote {\bibinfo {title}
  {Vortex loop dynamics and dynamical quantum phase transitions in 3d fermion
  matter},}\ }\href@noop {} {\  (\bibinfo {year} {2024})},\ \Eprint
  {http://arxiv.org/abs/2307.02985} {arXiv:2307.02985 [cond-mat.stat-mech]}
  \BibitemShut {NoStop}%
\bibitem [{\citenamefont {Abeling}\ and\ \citenamefont
  {Kehrein}(2016)}]{Abeling2016quantum}%
  \BibitemOpen
  \bibfield  {author} {\bibinfo {author} {\bibfnamefont {Nils~O.}\ \bibnamefont
  {Abeling}}\ and\ \bibinfo {author} {\bibfnamefont {Stefan}\ \bibnamefont
  {Kehrein}},\ }\bibfield  {title} {\enquote {\bibinfo {title} {Quantum quench
  dynamics in the transverse field ising model at nonzero temperatures},}\
  }\href {\doibase 10.1103/PhysRevB.93.104302} {\bibfield  {journal} {\bibinfo
  {journal} {Phys. Rev. B}\ }\textbf {\bibinfo {volume} {93}},\ \bibinfo
  {pages} {104302} (\bibinfo {year} {2016})}\BibitemShut {NoStop}%
\bibitem [{\citenamefont {Bhattacharya}\ \emph {et~al.}(2017)\citenamefont
  {Bhattacharya}, \citenamefont {Bandyopadhyay},\ and\ \citenamefont
  {Dutta}}]{Bhattacharya2017mixed}%
  \BibitemOpen
  \bibfield  {author} {\bibinfo {author} {\bibfnamefont {Utso}\ \bibnamefont
  {Bhattacharya}}, \bibinfo {author} {\bibfnamefont {Souvik}\ \bibnamefont
  {Bandyopadhyay}}, \ and\ \bibinfo {author} {\bibfnamefont {Amit}\
  \bibnamefont {Dutta}},\ }\bibfield  {title} {\enquote {\bibinfo {title}
  {Mixed state dynamical quantum phase transitions},}\ }\href {\doibase
  10.1103/PhysRevB.96.180303} {\bibfield  {journal} {\bibinfo  {journal} {Phys.
  Rev. B}\ }\textbf {\bibinfo {volume} {96}},\ \bibinfo {pages} {180303}
  (\bibinfo {year} {2017})}\BibitemShut {NoStop}%
\bibitem [{\citenamefont {Heyl}\ and\ \citenamefont
  {Budich}(2017)}]{Heyl2017dynamical}%
  \BibitemOpen
  \bibfield  {author} {\bibinfo {author} {\bibfnamefont {M.}~\bibnamefont
  {Heyl}}\ and\ \bibinfo {author} {\bibfnamefont {J.~C.}\ \bibnamefont
  {Budich}},\ }\bibfield  {title} {\enquote {\bibinfo {title} {Dynamical
  topological quantum phase transitions for mixed states},}\ }\href {\doibase
  10.1103/PhysRevB.96.180304} {\bibfield  {journal} {\bibinfo  {journal} {Phys.
  Rev. B}\ }\textbf {\bibinfo {volume} {96}},\ \bibinfo {pages} {180304}
  (\bibinfo {year} {2017})}\BibitemShut {NoStop}%
\bibitem [{\citenamefont {Sedlmayr}\ \emph
  {et~al.}(2018{\natexlab{b}})\citenamefont {Sedlmayr}, \citenamefont
  {Fleischhauer},\ and\ \citenamefont {Sirker}}]{Sedlmayr2018fate}%
  \BibitemOpen
  \bibfield  {author} {\bibinfo {author} {\bibfnamefont {N.}~\bibnamefont
  {Sedlmayr}}, \bibinfo {author} {\bibfnamefont {M.}~\bibnamefont
  {Fleischhauer}}, \ and\ \bibinfo {author} {\bibfnamefont {J.}~\bibnamefont
  {Sirker}},\ }\bibfield  {title} {\enquote {\bibinfo {title} {Fate of
  dynamical phase transitions at finite temperatures and in open systems},}\
  }\href {\doibase 10.1103/PhysRevB.97.045147} {\bibfield  {journal} {\bibinfo
  {journal} {Phys. Rev. B}\ }\textbf {\bibinfo {volume} {97}},\ \bibinfo
  {pages} {045147} (\bibinfo {year} {2018}{\natexlab{b}})}\BibitemShut
  {NoStop}%
\bibitem [{\citenamefont {Zache}\ \emph {et~al.}(2019)\citenamefont {Zache},
  \citenamefont {Mueller}, \citenamefont {Schneider}, \citenamefont
  {Jendrzejewski}, \citenamefont {Berges},\ and\ \citenamefont
  {Hauke}}]{Zache2019}%
  \BibitemOpen
  \bibfield  {author} {\bibinfo {author} {\bibfnamefont {T.~V.}\ \bibnamefont
  {Zache}}, \bibinfo {author} {\bibfnamefont {N.}~\bibnamefont {Mueller}},
  \bibinfo {author} {\bibfnamefont {J.~T.}\ \bibnamefont {Schneider}}, \bibinfo
  {author} {\bibfnamefont {F.}~\bibnamefont {Jendrzejewski}}, \bibinfo {author}
  {\bibfnamefont {J.}~\bibnamefont {Berges}}, \ and\ \bibinfo {author}
  {\bibfnamefont {P.}~\bibnamefont {Hauke}},\ }\bibfield  {title} {\enquote
  {\bibinfo {title} {Dynamical topological transitions in the massive schwinger
  model with a $\ensuremath{\theta}$ term},}\ }\href {\doibase
  10.1103/PhysRevLett.122.050403} {\bibfield  {journal} {\bibinfo  {journal}
  {Phys. Rev. Lett.}\ }\textbf {\bibinfo {volume} {122}},\ \bibinfo {pages}
  {050403} (\bibinfo {year} {2019})}\BibitemShut {NoStop}%
\bibitem [{\citenamefont {Huang}\ \emph {et~al.}(2019)\citenamefont {Huang},
  \citenamefont {Banerjee},\ and\ \citenamefont {Heyl}}]{Huang2019dynamical}%
  \BibitemOpen
  \bibfield  {author} {\bibinfo {author} {\bibfnamefont {Yi-Ping}\ \bibnamefont
  {Huang}}, \bibinfo {author} {\bibfnamefont {Debasish}\ \bibnamefont
  {Banerjee}}, \ and\ \bibinfo {author} {\bibfnamefont {Markus}\ \bibnamefont
  {Heyl}},\ }\bibfield  {title} {\enquote {\bibinfo {title} {Dynamical quantum
  phase transitions in u(1) quantum link models},}\ }\href {\doibase
  10.1103/PhysRevLett.122.250401} {\bibfield  {journal} {\bibinfo  {journal}
  {Phys. Rev. Lett.}\ }\textbf {\bibinfo {volume} {122}},\ \bibinfo {pages}
  {250401} (\bibinfo {year} {2019})}\BibitemShut {NoStop}%
\bibitem [{\citenamefont {Pedersen}\ and\ \citenamefont
  {Zinner}(2021)}]{Pedersen2021}%
  \BibitemOpen
  \bibfield  {author} {\bibinfo {author} {\bibfnamefont {Simon~Panyella}\
  \bibnamefont {Pedersen}}\ and\ \bibinfo {author} {\bibfnamefont
  {Nikolaj~Thomas}\ \bibnamefont {Zinner}},\ }\bibfield  {title} {\enquote
  {\bibinfo {title} {Lattice gauge theory and dynamical quantum phase
  transitions using noisy intermediate-scale quantum devices},}\ }\href
  {\doibase 10.1103/PhysRevB.103.235103} {\bibfield  {journal} {\bibinfo
  {journal} {Phys. Rev. B}\ }\textbf {\bibinfo {volume} {103}},\ \bibinfo
  {pages} {235103} (\bibinfo {year} {2021})}\BibitemShut {NoStop}%
\bibitem [{\citenamefont {Jensen}\ \emph {et~al.}(2022)\citenamefont {Jensen},
  \citenamefont {Pedersen},\ and\ \citenamefont {Zinner}}]{Jensen2022}%
  \BibitemOpen
  \bibfield  {author} {\bibinfo {author} {\bibfnamefont {Rasmus~Berg}\
  \bibnamefont {Jensen}}, \bibinfo {author} {\bibfnamefont {Simon~Panyella}\
  \bibnamefont {Pedersen}}, \ and\ \bibinfo {author} {\bibfnamefont
  {Nikolaj~Thomas}\ \bibnamefont {Zinner}},\ }\bibfield  {title} {\enquote
  {\bibinfo {title} {Dynamical quantum phase transitions in a noisy lattice
  gauge theory},}\ }\href {\doibase 10.1103/PhysRevB.105.224309} {\bibfield
  {journal} {\bibinfo  {journal} {Phys. Rev. B}\ }\textbf {\bibinfo {volume}
  {105}},\ \bibinfo {pages} {224309} (\bibinfo {year} {2022})}\BibitemShut
  {NoStop}%
\bibitem [{\citenamefont {Halimeh}\ \emph {et~al.}(2022)\citenamefont
  {Halimeh}, \citenamefont {Damme}, \citenamefont {Zache}, \citenamefont
  {Banerjee},\ and\ \citenamefont {Hauke}}]{Halimeh2021achieving}%
  \BibitemOpen
  \bibfield  {author} {\bibinfo {author} {\bibfnamefont {Jad~C.}\ \bibnamefont
  {Halimeh}}, \bibinfo {author} {\bibfnamefont {Maarten~Van}\ \bibnamefont
  {Damme}}, \bibinfo {author} {\bibfnamefont {Torsten~V.}\ \bibnamefont
  {Zache}}, \bibinfo {author} {\bibfnamefont {Debasish}\ \bibnamefont
  {Banerjee}}, \ and\ \bibinfo {author} {\bibfnamefont {Philipp}\ \bibnamefont
  {Hauke}},\ }\bibfield  {title} {\enquote {\bibinfo {title} {Achieving the
  quantum field theory limit in far-from-equilibrium quantum link models},}\
  }\href {\doibase 10.22331/q-2022-12-19-878} {\bibfield  {journal} {\bibinfo
  {journal} {{Quantum}}\ }\textbf {\bibinfo {volume} {6}},\ \bibinfo {pages}
  {878} (\bibinfo {year} {2022})}\BibitemShut {NoStop}%
\bibitem [{\citenamefont {Van~Damme}\ \emph {et~al.}(2022)\citenamefont
  {Van~Damme}, \citenamefont {Zache}, \citenamefont {Banerjee}, \citenamefont
  {Hauke},\ and\ \citenamefont {Halimeh}}]{VanDamme2022dynamical}%
  \BibitemOpen
  \bibfield  {author} {\bibinfo {author} {\bibfnamefont {Maarten}\ \bibnamefont
  {Van~Damme}}, \bibinfo {author} {\bibfnamefont {Torsten~V.}\ \bibnamefont
  {Zache}}, \bibinfo {author} {\bibfnamefont {Debasish}\ \bibnamefont
  {Banerjee}}, \bibinfo {author} {\bibfnamefont {Philipp}\ \bibnamefont
  {Hauke}}, \ and\ \bibinfo {author} {\bibfnamefont {Jad~C.}\ \bibnamefont
  {Halimeh}},\ }\bibfield  {title} {\enquote {\bibinfo {title} {Dynamical
  quantum phase transitions in spin-$s u(1)$ quantum link models},}\ }\href
  {\doibase 10.1103/PhysRevB.106.245110} {\bibfield  {journal} {\bibinfo
  {journal} {Phys. Rev. B}\ }\textbf {\bibinfo {volume} {106}},\ \bibinfo
  {pages} {245110} (\bibinfo {year} {2022})}\BibitemShut {NoStop}%
\bibitem [{\citenamefont {Mueller}\ \emph {et~al.}(2023)\citenamefont
  {Mueller}, \citenamefont {Carolan}, \citenamefont {Connelly}, \citenamefont
  {Davoudi}, \citenamefont {Dumitrescu},\ and\ \citenamefont
  {Yeter-Aydeniz}}]{Mueller2023quantum}%
  \BibitemOpen
  \bibfield  {author} {\bibinfo {author} {\bibfnamefont {Niklas}\ \bibnamefont
  {Mueller}}, \bibinfo {author} {\bibfnamefont {Joseph~A.}\ \bibnamefont
  {Carolan}}, \bibinfo {author} {\bibfnamefont {Andrew}\ \bibnamefont
  {Connelly}}, \bibinfo {author} {\bibfnamefont {Zohreh}\ \bibnamefont
  {Davoudi}}, \bibinfo {author} {\bibfnamefont {Eugene~F.}\ \bibnamefont
  {Dumitrescu}}, \ and\ \bibinfo {author} {\bibfnamefont {Kübra}\ \bibnamefont
  {Yeter-Aydeniz}},\ }\bibfield  {title} {\enquote {\bibinfo {title} {Quantum
  computation of dynamical quantum phase transitions and entanglement
  tomography in a lattice gauge theory},}\ }\href@noop {} {\  (\bibinfo {year}
  {2023})},\ \Eprint {http://arxiv.org/abs/2210.03089} {arXiv:2210.03089
  [quant-ph]} \BibitemShut {NoStop}%
\bibitem [{\citenamefont {Pomarico}\ \emph {et~al.}(2023)\citenamefont
  {Pomarico}, \citenamefont {Cosmai}, \citenamefont {Facchi}, \citenamefont
  {Lupo}, \citenamefont {Pascazio},\ and\ \citenamefont
  {Pepe}}]{Pomarico2023dynamical}%
  \BibitemOpen
  \bibfield  {author} {\bibinfo {author} {\bibfnamefont {Domenico}\
  \bibnamefont {Pomarico}}, \bibinfo {author} {\bibfnamefont {Leonardo}\
  \bibnamefont {Cosmai}}, \bibinfo {author} {\bibfnamefont {Paolo}\
  \bibnamefont {Facchi}}, \bibinfo {author} {\bibfnamefont {Cosmo}\
  \bibnamefont {Lupo}}, \bibinfo {author} {\bibfnamefont {Saverio}\
  \bibnamefont {Pascazio}}, \ and\ \bibinfo {author} {\bibfnamefont
  {Francesco~V.}\ \bibnamefont {Pepe}},\ }\bibfield  {title} {\enquote
  {\bibinfo {title} {Dynamical quantum phase transitions of the schwinger
  model: Real-time dynamics on ibm quantum},}\ }\href {\doibase
  10.3390/e25040608} {\bibfield  {journal} {\bibinfo  {journal} {Entropy}\
  }\textbf {\bibinfo {volume} {25}} (\bibinfo {year} {2023}),\
  10.3390/e25040608}\BibitemShut {NoStop}%
\bibitem [{\citenamefont {Van~Damme}\ \emph {et~al.}(2023)\citenamefont
  {Van~Damme}, \citenamefont {Desaules}, \citenamefont
  {Papi\ifmmode~\acute{c}\else \'{c}\fi{}},\ and\ \citenamefont
  {Halimeh}}]{VanDamme2023Anatomy}%
  \BibitemOpen
  \bibfield  {author} {\bibinfo {author} {\bibfnamefont {Maarten}\ \bibnamefont
  {Van~Damme}}, \bibinfo {author} {\bibfnamefont {Jean-Yves}\ \bibnamefont
  {Desaules}}, \bibinfo {author} {\bibfnamefont {Zlatko}\ \bibnamefont
  {Papi\ifmmode~\acute{c}\else \'{c}\fi{}}}, \ and\ \bibinfo {author}
  {\bibfnamefont {Jad~C.}\ \bibnamefont {Halimeh}},\ }\bibfield  {title}
  {\enquote {\bibinfo {title} {Anatomy of dynamical quantum phase
  transitions},}\ }\href {\doibase 10.1103/PhysRevResearch.5.033090} {\bibfield
   {journal} {\bibinfo  {journal} {Phys. Rev. Res.}\ }\textbf {\bibinfo
  {volume} {5}},\ \bibinfo {pages} {033090} (\bibinfo {year}
  {2023})}\BibitemShut {NoStop}%
\bibitem [{\citenamefont {Osborne}\ \emph {et~al.}(2023)\citenamefont
  {Osborne}, \citenamefont {McCulloch},\ and\ \citenamefont
  {Halimeh}}]{Osborne2023probing}%
  \BibitemOpen
  \bibfield  {author} {\bibinfo {author} {\bibfnamefont {Jesse}\ \bibnamefont
  {Osborne}}, \bibinfo {author} {\bibfnamefont {Ian~P.}\ \bibnamefont
  {McCulloch}}, \ and\ \bibinfo {author} {\bibfnamefont {Jad~C.}\ \bibnamefont
  {Halimeh}},\ }\bibfield  {title} {\enquote {\bibinfo {title} {Probing
  confinement through dynamical quantum phase transitions: From quantum spin
  models to lattice gauge theories},}\ }\href@noop {} {\  (\bibinfo {year}
  {2023})},\ \Eprint {http://arxiv.org/abs/2310.12210} {arXiv:2310.12210
  [cond-mat.quant-gas]} \BibitemShut {NoStop}%
\bibitem [{\citenamefont {Zhou}\ \emph {et~al.}(2018)\citenamefont {Zhou},
  \citenamefont {Wang}, \citenamefont {Wang},\ and\ \citenamefont
  {Gong}}]{Zhou2018Dynamical}%
  \BibitemOpen
  \bibfield  {author} {\bibinfo {author} {\bibfnamefont {Longwen}\ \bibnamefont
  {Zhou}}, \bibinfo {author} {\bibfnamefont {Qing-hai}\ \bibnamefont {Wang}},
  \bibinfo {author} {\bibfnamefont {Hailong}\ \bibnamefont {Wang}}, \ and\
  \bibinfo {author} {\bibfnamefont {Jiangbin}\ \bibnamefont {Gong}},\
  }\bibfield  {title} {\enquote {\bibinfo {title} {Dynamical quantum phase
  transitions in non-hermitian lattices},}\ }\href {\doibase
  10.1103/PhysRevA.98.022129} {\bibfield  {journal} {\bibinfo  {journal} {Phys.
  Rev. A}\ }\textbf {\bibinfo {volume} {98}},\ \bibinfo {pages} {022129}
  (\bibinfo {year} {2018})}\BibitemShut {NoStop}%
\bibitem [{\citenamefont {Zhou}\ and\ \citenamefont
  {Du}(2021)}]{Zhou2021Non-Hermitian}%
  \BibitemOpen
  \bibfield  {author} {\bibinfo {author} {\bibfnamefont {Longwen}\ \bibnamefont
  {Zhou}}\ and\ \bibinfo {author} {\bibfnamefont {Qianqian}\ \bibnamefont
  {Du}},\ }\bibfield  {title} {\enquote {\bibinfo {title} {Non-hermitian
  topological phases and dynamical quantum phase transitions: a generic
  connection},}\ }\href {\doibase 10.1088/1367-2630/ac0574} {\bibfield
  {journal} {\bibinfo  {journal} {New Journal of Physics}\ }\textbf {\bibinfo
  {volume} {23}},\ \bibinfo {pages} {063041} (\bibinfo {year}
  {2021})}\BibitemShut {NoStop}%
\bibitem [{\citenamefont {Hamazaki}(2021)}]{Hamazaki2021Exceptional}%
  \BibitemOpen
  \bibfield  {author} {\bibinfo {author} {\bibfnamefont {Ryusuke}\ \bibnamefont
  {Hamazaki}},\ }\bibfield  {title} {\enquote {\bibinfo {title} {Exceptional
  dynamical quantum phase transitions in periodically driven systems},}\ }\href
  {\doibase 10.1038/s41467-021-25355-3} {\bibfield  {journal} {\bibinfo
  {journal} {Nature Communications}\ }\textbf {\bibinfo {volume} {12}},\
  \bibinfo {pages} {5108} (\bibinfo {year} {2021})}\BibitemShut {NoStop}%
\bibitem [{\citenamefont {Mondal}\ and\ \citenamefont
  {Nag}(2022)}]{Nag2022Anomaly}%
  \BibitemOpen
  \bibfield  {author} {\bibinfo {author} {\bibfnamefont {Debashish}\
  \bibnamefont {Mondal}}\ and\ \bibinfo {author} {\bibfnamefont {Tanay}\
  \bibnamefont {Nag}},\ }\bibfield  {title} {\enquote {\bibinfo {title}
  {Anomaly in the dynamical quantum phase transition in a non-hermitian system
  with extended gapless phases},}\ }\href {\doibase
  10.1103/PhysRevB.106.054308} {\bibfield  {journal} {\bibinfo  {journal}
  {Phys. Rev. B}\ }\textbf {\bibinfo {volume} {106}},\ \bibinfo {pages}
  {054308} (\bibinfo {year} {2022})}\BibitemShut {NoStop}%
\bibitem [{\citenamefont {Mondal}\ and\ \citenamefont
  {Nag}(2023)}]{Nag2023Finite}%
  \BibitemOpen
  \bibfield  {author} {\bibinfo {author} {\bibfnamefont {Debashish}\
  \bibnamefont {Mondal}}\ and\ \bibinfo {author} {\bibfnamefont {Tanay}\
  \bibnamefont {Nag}},\ }\bibfield  {title} {\enquote {\bibinfo {title}
  {Finite-temperature dynamical quantum phase transition in a non-hermitian
  system},}\ }\href {\doibase 10.1103/PhysRevB.107.184311} {\bibfield
  {journal} {\bibinfo  {journal} {Phys. Rev. B}\ }\textbf {\bibinfo {volume}
  {107}},\ \bibinfo {pages} {184311} (\bibinfo {year} {2023})}\BibitemShut
  {NoStop}%
\bibitem [{\citenamefont {Mondal}\ and\ \citenamefont
  {Nag}(2024)}]{mondal2024persistent}%
  \BibitemOpen
  \bibfield  {author} {\bibinfo {author} {\bibfnamefont {Debashish}\
  \bibnamefont {Mondal}}\ and\ \bibinfo {author} {\bibfnamefont {Tanay}\
  \bibnamefont {Nag}},\ }\bibfield  {title} {\enquote {\bibinfo {title}
  {Persistent anomaly in dynamical quantum phase transition in long-range
  non-hermitian $p$-wave kitaev chian},}\ }\href@noop {} {\  (\bibinfo {year}
  {2024})},\ \Eprint {http://arxiv.org/abs/2402.04603} {arXiv:2402.04603
  [cond-mat.mes-hall]} \BibitemShut {NoStop}%
\bibitem [{\citenamefont {Kosior}\ and\ \citenamefont
  {Sacha}(2018)}]{Kosior2018a}%
  \BibitemOpen
  \bibfield  {author} {\bibinfo {author} {\bibfnamefont {Arkadiusz}\
  \bibnamefont {Kosior}}\ and\ \bibinfo {author} {\bibfnamefont {Krzysztof}\
  \bibnamefont {Sacha}},\ }\bibfield  {title} {\enquote {\bibinfo {title}
  {Dynamical quantum phase transitions in discrete time crystals},}\ }\href
  {\doibase 10.1103/PhysRevA.97.053621} {\bibfield  {journal} {\bibinfo
  {journal} {Phys. Rev. A}\ }\textbf {\bibinfo {volume} {97}},\ \bibinfo
  {pages} {053621} (\bibinfo {year} {2018})}\BibitemShut {NoStop}%
\bibitem [{\citenamefont {Kosior}\ \emph {et~al.}(2018)\citenamefont {Kosior},
  \citenamefont {Syrwid},\ and\ \citenamefont {Sacha}}]{Kosior2018b}%
  \BibitemOpen
  \bibfield  {author} {\bibinfo {author} {\bibfnamefont {Arkadiusz}\
  \bibnamefont {Kosior}}, \bibinfo {author} {\bibfnamefont {Andrzej}\
  \bibnamefont {Syrwid}}, \ and\ \bibinfo {author} {\bibfnamefont {Krzysztof}\
  \bibnamefont {Sacha}},\ }\bibfield  {title} {\enquote {\bibinfo {title}
  {Dynamical quantum phase transitions in systems with broken continuous time
  and space translation symmetries},}\ }\href {\doibase
  10.1103/PhysRevA.98.023612} {\bibfield  {journal} {\bibinfo  {journal} {Phys.
  Rev. A}\ }\textbf {\bibinfo {volume} {98}},\ \bibinfo {pages} {023612}
  (\bibinfo {year} {2018})}\BibitemShut {NoStop}%
\bibitem [{\citenamefont {Halimeh}\ \emph {et~al.}(2019)\citenamefont
  {Halimeh}, \citenamefont {Yegovtsev},\ and\ \citenamefont
  {Gurarie}}]{Halimeh2019dynamical}%
  \BibitemOpen
  \bibfield  {author} {\bibinfo {author} {\bibfnamefont {Jad~C.}\ \bibnamefont
  {Halimeh}}, \bibinfo {author} {\bibfnamefont {Nikolay}\ \bibnamefont
  {Yegovtsev}}, \ and\ \bibinfo {author} {\bibfnamefont {Victor}\ \bibnamefont
  {Gurarie}},\ }\bibfield  {title} {\enquote {\bibinfo {title} {Dynamical
  quantum phase transitions in many-body localized systems},}\ }\href@noop {}
  {\  (\bibinfo {year} {2019})},\ \Eprint {http://arxiv.org/abs/1903.03109}
  {arXiv:1903.03109 [cond-mat.stat-mech]} \BibitemShut {NoStop}%
\bibitem [{\citenamefont {Trapin}\ \emph {et~al.}(2021)\citenamefont {Trapin},
  \citenamefont {Halimeh},\ and\ \citenamefont {Heyl}}]{Trapin2021}%
  \BibitemOpen
  \bibfield  {author} {\bibinfo {author} {\bibfnamefont {Daniele}\ \bibnamefont
  {Trapin}}, \bibinfo {author} {\bibfnamefont {Jad~C.}\ \bibnamefont
  {Halimeh}}, \ and\ \bibinfo {author} {\bibfnamefont {Markus}\ \bibnamefont
  {Heyl}},\ }\bibfield  {title} {\enquote {\bibinfo {title} {Unconventional
  critical exponents at dynamical quantum phase transitions in a random ising
  chain},}\ }\href {\doibase 10.1103/PhysRevB.104.115159} {\bibfield  {journal}
  {\bibinfo  {journal} {Phys. Rev. B}\ }\textbf {\bibinfo {volume} {104}},\
  \bibinfo {pages} {115159} (\bibinfo {year} {2021})}\BibitemShut {NoStop}%
\bibitem [{\citenamefont {Jurcevic}\ \emph {et~al.}(2017)\citenamefont
  {Jurcevic}, \citenamefont {Shen}, \citenamefont {Hauke}, \citenamefont
  {Maier}, \citenamefont {Brydges}, \citenamefont {Hempel}, \citenamefont
  {Lanyon}, \citenamefont {Heyl}, \citenamefont {Blatt},\ and\ \citenamefont
  {Roos}}]{Jurcevic2017}%
  \BibitemOpen
  \bibfield  {author} {\bibinfo {author} {\bibfnamefont {P.}~\bibnamefont
  {Jurcevic}}, \bibinfo {author} {\bibfnamefont {H.}~\bibnamefont {Shen}},
  \bibinfo {author} {\bibfnamefont {P.}~\bibnamefont {Hauke}}, \bibinfo
  {author} {\bibfnamefont {C.}~\bibnamefont {Maier}}, \bibinfo {author}
  {\bibfnamefont {T.}~\bibnamefont {Brydges}}, \bibinfo {author} {\bibfnamefont
  {C.}~\bibnamefont {Hempel}}, \bibinfo {author} {\bibfnamefont {B.~P.}\
  \bibnamefont {Lanyon}}, \bibinfo {author} {\bibfnamefont {M.}~\bibnamefont
  {Heyl}}, \bibinfo {author} {\bibfnamefont {R.}~\bibnamefont {Blatt}}, \ and\
  \bibinfo {author} {\bibfnamefont {C.~F.}\ \bibnamefont {Roos}},\ }\bibfield
  {title} {\enquote {\bibinfo {title} {Direct observation of dynamical quantum
  phase transitions in an interacting many-body system},}\ }\href {\doibase
  10.1103/PhysRevLett.119.080501} {\bibfield  {journal} {\bibinfo  {journal}
  {Phys. Rev. Lett.}\ }\textbf {\bibinfo {volume} {119}},\ \bibinfo {pages}
  {080501} (\bibinfo {year} {2017})}\BibitemShut {NoStop}%
\bibitem [{\citenamefont {Fl{\"a}schner}\ \emph {et~al.}(2018)\citenamefont
  {Fl{\"a}schner}, \citenamefont {Vogel}, \citenamefont {Tarnowski},
  \citenamefont {Rem}, \citenamefont {L{\"u}hmann}, \citenamefont {Heyl},
  \citenamefont {Budich}, \citenamefont {Mathey}, \citenamefont {Sengstock},\
  and\ \citenamefont {Weitenberg}}]{Flaeschner2018}%
  \BibitemOpen
  \bibfield  {author} {\bibinfo {author} {\bibfnamefont {N.}~\bibnamefont
  {Fl{\"a}schner}}, \bibinfo {author} {\bibfnamefont {D.}~\bibnamefont
  {Vogel}}, \bibinfo {author} {\bibfnamefont {M.}~\bibnamefont {Tarnowski}},
  \bibinfo {author} {\bibfnamefont {B.~S.}\ \bibnamefont {Rem}}, \bibinfo
  {author} {\bibfnamefont {D.-S.}\ \bibnamefont {L{\"u}hmann}}, \bibinfo
  {author} {\bibfnamefont {M.}~\bibnamefont {Heyl}}, \bibinfo {author}
  {\bibfnamefont {J.~C.}\ \bibnamefont {Budich}}, \bibinfo {author}
  {\bibfnamefont {L.}~\bibnamefont {Mathey}}, \bibinfo {author} {\bibfnamefont
  {K.}~\bibnamefont {Sengstock}}, \ and\ \bibinfo {author} {\bibfnamefont
  {C.}~\bibnamefont {Weitenberg}},\ }\bibfield  {title} {\enquote {\bibinfo
  {title} {Observation of dynamical vortices after quenches in a system with
  topology},}\ }\href {https://doi.org/10.1038/s41567-017-0013-8} {\bibfield
  {journal} {\bibinfo  {journal} {Nature Physics}\ }\textbf {\bibinfo {volume}
  {14}},\ \bibinfo {pages} {265--268} (\bibinfo {year} {2018})}\BibitemShut
  {NoStop}%
\bibitem [{\citenamefont {Nie}\ \emph {et~al.}(2020)\citenamefont {Nie},
  \citenamefont {Wei}, \citenamefont {Chen}, \citenamefont {Zhang},
  \citenamefont {Zhao}, \citenamefont {Qiu}, \citenamefont {Tian},
  \citenamefont {Ji}, \citenamefont {Xin}, \citenamefont {Lu},\ and\
  \citenamefont {Li}}]{Nie2020experimental}%
  \BibitemOpen
  \bibfield  {author} {\bibinfo {author} {\bibfnamefont {Xinfang}\ \bibnamefont
  {Nie}}, \bibinfo {author} {\bibfnamefont {Bo-Bo}\ \bibnamefont {Wei}},
  \bibinfo {author} {\bibfnamefont {Xi}~\bibnamefont {Chen}}, \bibinfo {author}
  {\bibfnamefont {Ze}~\bibnamefont {Zhang}}, \bibinfo {author} {\bibfnamefont
  {Xiuzhu}\ \bibnamefont {Zhao}}, \bibinfo {author} {\bibfnamefont {Chudan}\
  \bibnamefont {Qiu}}, \bibinfo {author} {\bibfnamefont {Yu}~\bibnamefont
  {Tian}}, \bibinfo {author} {\bibfnamefont {Yunlan}\ \bibnamefont {Ji}},
  \bibinfo {author} {\bibfnamefont {Tao}\ \bibnamefont {Xin}}, \bibinfo
  {author} {\bibfnamefont {Dawei}\ \bibnamefont {Lu}}, \ and\ \bibinfo {author}
  {\bibfnamefont {Jun}\ \bibnamefont {Li}},\ }\bibfield  {title} {\enquote
  {\bibinfo {title} {Experimental observation of equilibrium and dynamical
  quantum phase transitions via out-of-time-ordered correlators},}\ }\href
  {\doibase 10.1103/PhysRevLett.124.250601} {\bibfield  {journal} {\bibinfo
  {journal} {Phys. Rev. Lett.}\ }\textbf {\bibinfo {volume} {124}},\ \bibinfo
  {pages} {250601} (\bibinfo {year} {2020})}\BibitemShut {NoStop}%
\bibitem [{\citenamefont {Caprio}\ \emph {et~al.}(2008)\citenamefont {Caprio},
  \citenamefont {Cejnar},\ and\ \citenamefont {Iachello}}]{Caprio2008}%
  \BibitemOpen
  \bibfield  {author} {\bibinfo {author} {\bibfnamefont {M.A.}\ \bibnamefont
  {Caprio}}, \bibinfo {author} {\bibfnamefont {P.}~\bibnamefont {Cejnar}}, \
  and\ \bibinfo {author} {\bibfnamefont {F.}~\bibnamefont {Iachello}},\
  }\bibfield  {title} {\enquote {\bibinfo {title} {Excited state quantum phase
  transitions in many-body systems},}\ }\href {\doibase
  10.1016/j.aop.2007.06.011} {\bibfield  {journal} {\bibinfo  {journal} {Annals
  of Physics}\ }\textbf {\bibinfo {volume} {323}},\ \bibinfo {pages}
  {1106--1135} (\bibinfo {year} {2008})}\BibitemShut {NoStop}%
\bibitem [{\citenamefont {Cejnar}\ \emph {et~al.}(2021)\citenamefont {Cejnar},
  \citenamefont {Stránský}, \citenamefont {Macek},\ and\ \citenamefont
  {Kloc}}]{Cejnar2021}%
  \BibitemOpen
  \bibfield  {author} {\bibinfo {author} {\bibfnamefont {Pavel}\ \bibnamefont
  {Cejnar}}, \bibinfo {author} {\bibfnamefont {Pavel}\ \bibnamefont
  {Stránský}}, \bibinfo {author} {\bibfnamefont {Michal}\ \bibnamefont
  {Macek}}, \ and\ \bibinfo {author} {\bibfnamefont {Michal}\ \bibnamefont
  {Kloc}},\ }\bibfield  {title} {\enquote {\bibinfo {title} {Excited-state
  quantum phase transitions},}\ }\href {\doibase 10.1088/1751-8121/abdfe8}
  {\bibfield  {journal} {\bibinfo  {journal} {Journal of Physics A:
  Mathematical and Theoretical}\ }\textbf {\bibinfo {volume} {54}},\ \bibinfo
  {pages} {133001} (\bibinfo {year} {2021})}\BibitemShut {NoStop}%
\bibitem [{\citenamefont {Relaño}\ \emph {et~al.}(2008)\citenamefont
  {Relaño}, \citenamefont {Arias}, \citenamefont {Dukelsky}, \citenamefont
  {García-Ramos},\ and\ \citenamefont {Pérez-Fernández}}]{Relano2008}%
  \BibitemOpen
  \bibfield  {author} {\bibinfo {author} {\bibfnamefont {A.}~\bibnamefont
  {Relaño}}, \bibinfo {author} {\bibfnamefont {J.~M.}\ \bibnamefont {Arias}},
  \bibinfo {author} {\bibfnamefont {J.}~\bibnamefont {Dukelsky}}, \bibinfo
  {author} {\bibfnamefont {J.~E.}\ \bibnamefont {García-Ramos}}, \ and\
  \bibinfo {author} {\bibfnamefont {P.}~\bibnamefont {Pérez-Fernández}},\
  }\bibfield  {title} {\enquote {\bibinfo {title} {Decoherence as a signature
  of an excited-state quantum phase transition},}\ }\href {\doibase
  10.1103/PhysRevA.78.060102} {\bibfield  {journal} {\bibinfo  {journal}
  {Physical Review A}\ }\textbf {\bibinfo {volume} {78}},\ \bibinfo {pages}
  {060102} (\bibinfo {year} {2008})}\BibitemShut {NoStop}%
\bibitem [{\citenamefont {Pérez-Fernández}\ \emph {et~al.}(2009)\citenamefont
  {Pérez-Fernández}, \citenamefont {Relaño}, \citenamefont {Arias},
  \citenamefont {Dukelsky},\ and\ \citenamefont
  {García-Ramos}}]{PerezFernandez2009}%
  \BibitemOpen
  \bibfield  {author} {\bibinfo {author} {\bibfnamefont {P.}~\bibnamefont
  {Pérez-Fernández}}, \bibinfo {author} {\bibfnamefont {A.}~\bibnamefont
  {Relaño}}, \bibinfo {author} {\bibfnamefont {J.~M.}\ \bibnamefont {Arias}},
  \bibinfo {author} {\bibfnamefont {J.}~\bibnamefont {Dukelsky}}, \ and\
  \bibinfo {author} {\bibfnamefont {J.~E.}\ \bibnamefont {García-Ramos}},\
  }\bibfield  {title} {\enquote {\bibinfo {title} {Decoherence due to an
  excited-state quantum phase transition in a two-level boson model},}\ }\href
  {\doibase 10.1103/PhysRevA.80.032111} {\bibfield  {journal} {\bibinfo
  {journal} {Physical Review A}\ }\textbf {\bibinfo {volume} {80}},\ \bibinfo
  {pages} {032111} (\bibinfo {year} {2009})}\BibitemShut {NoStop}%
\bibitem [{\citenamefont {Pérez-Fernández}\ \emph {et~al.}(2011)\citenamefont
  {Pérez-Fernández}, \citenamefont {Cejnar}, \citenamefont {Arias},
  \citenamefont {Dukelsky}, \citenamefont {García-Ramos},\ and\ \citenamefont
  {Relaño}}]{PerezFernandez2011}%
  \BibitemOpen
  \bibfield  {author} {\bibinfo {author} {\bibfnamefont {P.}~\bibnamefont
  {Pérez-Fernández}}, \bibinfo {author} {\bibfnamefont {P.}~\bibnamefont
  {Cejnar}}, \bibinfo {author} {\bibfnamefont {J.~M.}\ \bibnamefont {Arias}},
  \bibinfo {author} {\bibfnamefont {J.}~\bibnamefont {Dukelsky}}, \bibinfo
  {author} {\bibfnamefont {J.~E.}\ \bibnamefont {García-Ramos}}, \ and\
  \bibinfo {author} {\bibfnamefont {A.}~\bibnamefont {Relaño}},\ }\bibfield
  {title} {\enquote {\bibinfo {title} {Quantum quench influenced by an
  excited-state phase transition},}\ }\href {\doibase
  10.1103/PhysRevA.83.033802} {\bibfield  {journal} {\bibinfo  {journal}
  {Physical Review A}\ }\textbf {\bibinfo {volume} {83}},\ \bibinfo {pages}
  {033802} (\bibinfo {year} {2011})}\BibitemShut {NoStop}%
\bibitem [{\citenamefont {Santos}\ and\ \citenamefont
  {Pérez-Bernal}(2015)}]{Santos2015}%
  \BibitemOpen
  \bibfield  {author} {\bibinfo {author} {\bibfnamefont {Lea~F.}\ \bibnamefont
  {Santos}}\ and\ \bibinfo {author} {\bibfnamefont {Francisco}\ \bibnamefont
  {Pérez-Bernal}},\ }\bibfield  {title} {\enquote {\bibinfo {title} {Structure
  of eigenstates and quench dynamics at an excited-state quantum phase
  transition},}\ }\href {\doibase 10.1103/PhysRevA.92.050101} {\bibfield
  {journal} {\bibinfo  {journal} {Physical Review A}\ }\textbf {\bibinfo
  {volume} {92}},\ \bibinfo {pages} {050101} (\bibinfo {year}
  {2015})}\BibitemShut {NoStop}%
\bibitem [{\citenamefont {Lóbez}\ and\ \citenamefont
  {Relaño}(2016)}]{Lobez2016}%
  \BibitemOpen
  \bibfield  {author} {\bibinfo {author} {\bibfnamefont {C.~M.}\ \bibnamefont
  {Lóbez}}\ and\ \bibinfo {author} {\bibfnamefont {A.}~\bibnamefont
  {Relaño}},\ }\bibfield  {title} {\enquote {\bibinfo {title} {Entropy, chaos,
  and excited-state quantum phase transitions in the dicke model},}\ }\href
  {\doibase 10.1103/PhysRevE.94.012140} {\bibfield  {journal} {\bibinfo
  {journal} {Physical Review E}\ }\textbf {\bibinfo {volume} {94}},\ \bibinfo
  {pages} {012140} (\bibinfo {year} {2016})}\BibitemShut {NoStop}%
\bibitem [{\citenamefont {Pérez‐Bernal}\ and\ \citenamefont
  {Santos}(2017)}]{PerezBernal2017}%
  \BibitemOpen
  \bibfield  {author} {\bibinfo {author} {\bibfnamefont {Francisco}\
  \bibnamefont {Pérez‐Bernal}}\ and\ \bibinfo {author} {\bibfnamefont
  {Lea~F.}\ \bibnamefont {Santos}},\ }\bibfield  {title} {\enquote {\bibinfo
  {title} {Effects of excited state quantum phase transitions on system
  dynamics},}\ }\href {\doibase 10.1002/prop.201600035} {\bibfield  {journal}
  {\bibinfo  {journal} {Fortschritte der Physik}\ }\textbf {\bibinfo {volume}
  {65}},\ \bibinfo {pages} {1600035} (\bibinfo {year} {2017})}\BibitemShut
  {NoStop}%
\bibitem [{\citenamefont {Kloc}\ \emph {et~al.}(2018)\citenamefont {Kloc},
  \citenamefont {Stránský},\ and\ \citenamefont {Cejnar}}]{Kloc2018}%
  \BibitemOpen
  \bibfield  {author} {\bibinfo {author} {\bibfnamefont {Michal}\ \bibnamefont
  {Kloc}}, \bibinfo {author} {\bibfnamefont {Pavel}\ \bibnamefont
  {Stránský}}, \ and\ \bibinfo {author} {\bibfnamefont {Pavel}\ \bibnamefont
  {Cejnar}},\ }\bibfield  {title} {\enquote {\bibinfo {title} {Quantum quench
  dynamics in dicke superradiance models},}\ }\href {\doibase
  10.1103/PhysRevA.98.013836} {\bibfield  {journal} {\bibinfo  {journal}
  {Physical Review A}\ }\textbf {\bibinfo {volume} {98}},\ \bibinfo {pages}
  {013836} (\bibinfo {year} {2018})}\BibitemShut {NoStop}%
\bibitem [{\citenamefont {Kopylov}\ and\ \citenamefont
  {Brandes}(2015)}]{Kopylov2015}%
  \BibitemOpen
  \bibfield  {author} {\bibinfo {author} {\bibfnamefont {Wassilij}\
  \bibnamefont {Kopylov}}\ and\ \bibinfo {author} {\bibfnamefont {Tobias}\
  \bibnamefont {Brandes}},\ }\bibfield  {title} {\enquote {\bibinfo {title}
  {Time delayed control of excited state quantum phase transitions in the
  lipkin–meshkov–glick model},}\ }\href {\doibase
  10.1088/1367-2630/17/10/103031} {\bibfield  {journal} {\bibinfo  {journal}
  {New Journal of Physics}\ }\textbf {\bibinfo {volume} {17}},\ \bibinfo
  {pages} {103031} (\bibinfo {year} {2015})}\BibitemShut {NoStop}%
\bibitem [{\citenamefont {Wang}\ and\ \citenamefont {Quan}(2017)}]{Wang2017}%
  \BibitemOpen
  \bibfield  {author} {\bibinfo {author} {\bibfnamefont {Qian}\ \bibnamefont
  {Wang}}\ and\ \bibinfo {author} {\bibfnamefont {H.~T.}\ \bibnamefont
  {Quan}},\ }\bibfield  {title} {\enquote {\bibinfo {title} {Probing the
  excited-state quantum phase transition through statistics of loschmidt echo
  and quantum work},}\ }\href {\doibase 10.1103/PhysRevE.96.032142} {\bibfield
  {journal} {\bibinfo  {journal} {Physical Review E}\ }\textbf {\bibinfo
  {volume} {96}},\ \bibinfo {pages} {032142} (\bibinfo {year}
  {2017})}\BibitemShut {NoStop}%
\bibitem [{\citenamefont {Puebla}\ \emph {et~al.}(2013)\citenamefont {Puebla},
  \citenamefont {Relaño},\ and\ \citenamefont {Retamosa}}]{Puebla2013}%
  \BibitemOpen
  \bibfield  {author} {\bibinfo {author} {\bibfnamefont {Ricardo}\ \bibnamefont
  {Puebla}}, \bibinfo {author} {\bibfnamefont {Armando}\ \bibnamefont
  {Relaño}}, \ and\ \bibinfo {author} {\bibfnamefont {Joaquín}\ \bibnamefont
  {Retamosa}},\ }\bibfield  {title} {\enquote {\bibinfo {title} {Excited-state
  phase transition leading to symmetry-breaking steady states in the dicke
  model},}\ }\href {\doibase 10.1103/PhysRevA.87.023819} {\bibfield  {journal}
  {\bibinfo  {journal} {Physical Review A}\ }\textbf {\bibinfo {volume} {87}},\
  \bibinfo {pages} {023819} (\bibinfo {year} {2013})}\BibitemShut {NoStop}%
\bibitem [{\citenamefont {Puebla}\ and\ \citenamefont
  {Relaño}(2014)}]{Puebla2014}%
  \BibitemOpen
  \bibfield  {author} {\bibinfo {author} {\bibfnamefont {Ricardo}\ \bibnamefont
  {Puebla}}\ and\ \bibinfo {author} {\bibfnamefont {Armando}\ \bibnamefont
  {Relaño}},\ }\bibfield  {title} {\enquote {\bibinfo {title} {Non-thermal
  excited-state quantum phase transitions},}\ }\href {\doibase
  10.1209/0295-5075/104/50007} {\bibfield  {journal} {\bibinfo  {journal} {EPL
  (Europhysics Letters)}\ }\textbf {\bibinfo {volume} {104}},\ \bibinfo {pages}
  {50007} (\bibinfo {year} {2014})}\BibitemShut {NoStop}%
\bibitem [{\citenamefont {Bastidas}\ \emph {et~al.}(2014)\citenamefont
  {Bastidas}, \citenamefont {Pérez-Fernández}, \citenamefont {Vogl},\ and\
  \citenamefont {Brandes}}]{Bastidas2014}%
  \BibitemOpen
  \bibfield  {author} {\bibinfo {author} {\bibfnamefont {Victor~Manuel}\
  \bibnamefont {Bastidas}}, \bibinfo {author} {\bibfnamefont {Pedro}\
  \bibnamefont {Pérez-Fernández}}, \bibinfo {author} {\bibfnamefont {Malte}\
  \bibnamefont {Vogl}}, \ and\ \bibinfo {author} {\bibfnamefont {Tobias}\
  \bibnamefont {Brandes}},\ }\bibfield  {title} {\enquote {\bibinfo {title}
  {Quantum criticality and dynamical instability in the kicked-top model},}\
  }\href {\doibase 10.1103/PhysRevLett.112.140408} {\bibfield  {journal}
  {\bibinfo  {journal} {Physical Review Letters}\ }\textbf {\bibinfo {volume}
  {112}},\ \bibinfo {pages} {140408} (\bibinfo {year} {2014})}\BibitemShut
  {NoStop}%
\bibitem [{\citenamefont {Puebla}\ and\ \citenamefont
  {Relaño}(2015)}]{Puebla2015}%
  \BibitemOpen
  \bibfield  {author} {\bibinfo {author} {\bibfnamefont {Ricardo}\ \bibnamefont
  {Puebla}}\ and\ \bibinfo {author} {\bibfnamefont {Armando}\ \bibnamefont
  {Relaño}},\ }\bibfield  {title} {\enquote {\bibinfo {title} {Irreversible
  processes without energy dissipation in an isolated {Lipkin-Meshkov-Glick}
  model},}\ }\href {\doibase 10.1103/PhysRevE.92.012101} {\bibfield  {journal}
  {\bibinfo  {journal} {Physical Review E}\ }\textbf {\bibinfo {volume} {92}},\
  \bibinfo {pages} {012101} (\bibinfo {year} {2015})}\BibitemShut {NoStop}%
\bibitem [{\citenamefont {Hummel}\ \emph {et~al.}(2019)\citenamefont {Hummel},
  \citenamefont {Geiger}, \citenamefont {Urbina},\ and\ \citenamefont
  {Richter}}]{Hummel2019}%
  \BibitemOpen
  \bibfield  {author} {\bibinfo {author} {\bibfnamefont {Quirin}\ \bibnamefont
  {Hummel}}, \bibinfo {author} {\bibfnamefont {Benjamin}\ \bibnamefont
  {Geiger}}, \bibinfo {author} {\bibfnamefont {Juan~Diego}\ \bibnamefont
  {Urbina}}, \ and\ \bibinfo {author} {\bibfnamefont {Klaus}\ \bibnamefont
  {Richter}},\ }\bibfield  {title} {\enquote {\bibinfo {title} {Reversible
  quantum information spreading in many-body systems near criticality},}\
  }\href {\doibase 10.1103/PhysRevLett.123.160401} {\bibfield  {journal}
  {\bibinfo  {journal} {Physical Review Letters}\ }\textbf {\bibinfo {volume}
  {123}},\ \bibinfo {pages} {160401} (\bibinfo {year} {2019})}\BibitemShut
  {NoStop}%
\bibitem [{\citenamefont {Ángel L.~Corps}\ and\ \citenamefont
  {Relaño}(2021)}]{Corps2021}%
  \BibitemOpen
  \bibfield  {author} {\bibinfo {author} {\bibnamefont {Ángel L.~Corps}}\ and\
  \bibinfo {author} {\bibfnamefont {Armando}\ \bibnamefont {Relaño}},\
  }\bibfield  {title} {\enquote {\bibinfo {title} {Constant of motion
  identifying excited-state quantum phases},}\ }\href {\doibase
  10.1103/PhysRevLett.127.130602} {\bibfield  {journal} {\bibinfo  {journal}
  {Physical Review Letters}\ }\textbf {\bibinfo {volume} {127}},\ \bibinfo
  {pages} {130602} (\bibinfo {year} {2021})}\BibitemShut {NoStop}%
\bibitem [{\citenamefont {Ángel L.~Corps}\ and\ \citenamefont
  {Relaño}(2022)}]{Corps2022}%
  \BibitemOpen
  \bibfield  {author} {\bibinfo {author} {\bibnamefont {Ángel L.~Corps}}\ and\
  \bibinfo {author} {\bibfnamefont {Armando}\ \bibnamefont {Relaño}},\
  }\bibfield  {title} {\enquote {\bibinfo {title} {Dynamical and excited-state
  quantum phase transitions in collective systems},}\ }\href {\doibase
  10.1103/PhysRevB.106.024311} {\bibfield  {journal} {\bibinfo  {journal}
  {Physical Review B}\ }\textbf {\bibinfo {volume} {106}},\ \bibinfo {pages}
  {024311} (\bibinfo {year} {2022})}\BibitemShut {NoStop}%
\bibitem [{\citenamefont {Ángel L.~Corps}\ and\ \citenamefont
  {Relaño}(2023)}]{Corps2023}%
  \BibitemOpen
  \bibfield  {author} {\bibinfo {author} {\bibnamefont {Ángel L.~Corps}}\ and\
  \bibinfo {author} {\bibfnamefont {Armando}\ \bibnamefont {Relaño}},\
  }\bibfield  {title} {\enquote {\bibinfo {title} {Theory of dynamical phase
  transitions in quantum systems with symmetry-breaking eigenstates},}\ }\href
  {\doibase 10.1103/PhysRevLett.130.100402} {\bibfield  {journal} {\bibinfo
  {journal} {Physical Review Letters}\ }\textbf {\bibinfo {volume} {130}},\
  \bibinfo {pages} {100402} (\bibinfo {year} {2023})}\BibitemShut {NoStop}%
\bibitem [{\citenamefont {Halimeh}\ \emph
  {et~al.}(2021{\natexlab{b}})\citenamefont {Halimeh}, \citenamefont {Trapin},
  \citenamefont {Van~Damme},\ and\ \citenamefont {Heyl}}]{Halimeh2021Local}%
  \BibitemOpen
  \bibfield  {author} {\bibinfo {author} {\bibfnamefont {Jad~C.}\ \bibnamefont
  {Halimeh}}, \bibinfo {author} {\bibfnamefont {Daniele}\ \bibnamefont
  {Trapin}}, \bibinfo {author} {\bibfnamefont {Maarten}\ \bibnamefont
  {Van~Damme}}, \ and\ \bibinfo {author} {\bibfnamefont {Markus}\ \bibnamefont
  {Heyl}},\ }\bibfield  {title} {\enquote {\bibinfo {title} {Local measures of
  dynamical quantum phase transitions},}\ }\href {\doibase
  10.1103/PhysRevB.104.075130} {\bibfield  {journal} {\bibinfo  {journal}
  {Phys. Rev. B}\ }\textbf {\bibinfo {volume} {104}},\ \bibinfo {pages}
  {075130} (\bibinfo {year} {2021}{\natexlab{b}})}\BibitemShut {NoStop}%
\bibitem [{\citenamefont {Wu}(2020)}]{Wu2020}%
  \BibitemOpen
  \bibfield  {author} {\bibinfo {author} {\bibfnamefont {Yantao}\ \bibnamefont
  {Wu}},\ }\bibfield  {title} {\enquote {\bibinfo {title} {Dynamical quantum
  phase transitions of quantum spin chains with a loschmidt-rate critical
  exponent equal to $\frac{1}{2}$},}\ }\href {\doibase
  10.1103/PhysRevB.101.064427} {\bibfield  {journal} {\bibinfo  {journal}
  {Phys. Rev. B}\ }\textbf {\bibinfo {volume} {101}},\ \bibinfo {pages}
  {064427} (\bibinfo {year} {2020})}\BibitemShut {NoStop}%
\bibitem [{\citenamefont {Lipkin}\ \emph {et~al.}(1965)\citenamefont {Lipkin},
  \citenamefont {Meshkov},\ and\ \citenamefont {Glick}}]{Lipkin1965}%
  \BibitemOpen
  \bibfield  {author} {\bibinfo {author} {\bibfnamefont {H.J.}\ \bibnamefont
  {Lipkin}}, \bibinfo {author} {\bibfnamefont {N.}~\bibnamefont {Meshkov}}, \
  and\ \bibinfo {author} {\bibfnamefont {A.J.}\ \bibnamefont {Glick}},\
  }\bibfield  {title} {\enquote {\bibinfo {title} {Validity of many-body
  approximation methods for a solvable model: (i). exact solutions and
  perturbation theory},}\ }\href {\doibase 10.1016/0029-5582(65)90862-X}
  {\bibfield  {journal} {\bibinfo  {journal} {Nuclear Physics}\ }\textbf
  {\bibinfo {volume} {62}},\ \bibinfo {pages} {188--198} (\bibinfo {year}
  {1965})}\BibitemShut {NoStop}%
\bibitem [{\citenamefont {Meshkov}\ \emph {et~al.}(1965)\citenamefont
  {Meshkov}, \citenamefont {Glick},\ and\ \citenamefont
  {Lipkin}}]{Meshkov1965}%
  \BibitemOpen
  \bibfield  {author} {\bibinfo {author} {\bibfnamefont {N.}~\bibnamefont
  {Meshkov}}, \bibinfo {author} {\bibfnamefont {A.J.}\ \bibnamefont {Glick}}, \
  and\ \bibinfo {author} {\bibfnamefont {H.J.}\ \bibnamefont {Lipkin}},\
  }\bibfield  {title} {\enquote {\bibinfo {title} {Validity of many-body
  approximation methods for a solvable model: (ii). linearization
  procedures},}\ }\href {\doibase 10.1016/0029-5582(65)90863-1} {\bibfield
  {journal} {\bibinfo  {journal} {Nuclear Physics}\ }\textbf {\bibinfo {volume}
  {62}},\ \bibinfo {pages} {199--210} (\bibinfo {year} {1965})}\BibitemShut
  {NoStop}%
\bibitem [{\citenamefont {Glick}\ \emph {et~al.}(1965)\citenamefont {Glick},
  \citenamefont {Lipkin},\ and\ \citenamefont {Meshkov}}]{Glick1965}%
  \BibitemOpen
  \bibfield  {author} {\bibinfo {author} {\bibfnamefont {A.J.}\ \bibnamefont
  {Glick}}, \bibinfo {author} {\bibfnamefont {H.J.}\ \bibnamefont {Lipkin}}, \
  and\ \bibinfo {author} {\bibfnamefont {N.}~\bibnamefont {Meshkov}},\
  }\bibfield  {title} {\enquote {\bibinfo {title} {Validity of many-body
  approximation methods for a solvable model: (iii). diagram summations},}\
  }\href {\doibase 10.1016/0029-5582(65)90864-3} {\bibfield  {journal}
  {\bibinfo  {journal} {Nuclear Physics}\ }\textbf {\bibinfo {volume} {62}},\
  \bibinfo {pages} {211--224} (\bibinfo {year} {1965})}\BibitemShut {NoStop}%
\bibitem [{\citenamefont {Pérez-Fernández}\ and\ \citenamefont
  {Relaño}(2017)}]{PerezFernandez2017dicke}%
  \BibitemOpen
  \bibfield  {author} {\bibinfo {author} {\bibfnamefont {P.}~\bibnamefont
  {Pérez-Fernández}}\ and\ \bibinfo {author} {\bibfnamefont {A.}~\bibnamefont
  {Relaño}},\ }\bibfield  {title} {\enquote {\bibinfo {title} {From thermal to
  excited-state quantum phase transition: The dicke model},}\ }\href {\doibase
  10.1103/PhysRevE.96.012121} {\bibfield  {journal} {\bibinfo  {journal}
  {Physical Review E}\ }\textbf {\bibinfo {volume} {96}},\ \bibinfo {pages}
  {012121} (\bibinfo {year} {2017})}\BibitemShut {NoStop}%
\bibitem [{\citenamefont {L.~Corps}\ and\ \citenamefont
  {Relaño}(2023)}]{Corps2023arxiv}%
  \BibitemOpen
  \bibfield  {author} {\bibinfo {author} {\bibfnamefont {Ángel}\ \bibnamefont
  {L.~Corps}}\ and\ \bibinfo {author} {\bibfnamefont {Armando}\ \bibnamefont
  {Relaño}},\ }\bibfield  {title} {\enquote {\bibinfo {title} {General theory
  for discrete symmetry-breaking equilibrium states},}\ }\href {\doibase
  10.48550/arXiv.2303.18020} {\  (\bibinfo {year} {2023}),\
  10.48550/arXiv.2303.18020}\BibitemShut {NoStop}%
\bibitem [{SM()}]{SM}%
  \BibitemOpen
  \href@noop {} {}\bibinfo {howpublished} {See Supplemental Material for
  supporting numerical results}\BibitemShut {NoStop}%
\bibitem [{\citenamefont {Halimeh}\ \emph {et~al.}(2020)\citenamefont
  {Halimeh}, \citenamefont {Van~Damme}, \citenamefont {Zauner-Stauber},\ and\
  \citenamefont {Vanderstraeten}}]{Halimeh2020quasiparticle}%
  \BibitemOpen
  \bibfield  {author} {\bibinfo {author} {\bibfnamefont {Jad~C.}\ \bibnamefont
  {Halimeh}}, \bibinfo {author} {\bibfnamefont {Maarten}\ \bibnamefont
  {Van~Damme}}, \bibinfo {author} {\bibfnamefont {Valentin}\ \bibnamefont
  {Zauner-Stauber}}, \ and\ \bibinfo {author} {\bibfnamefont {Laurens}\
  \bibnamefont {Vanderstraeten}},\ }\bibfield  {title} {\enquote {\bibinfo
  {title} {Quasiparticle origin of dynamical quantum phase transitions},}\
  }\href {\doibase 10.1103/PhysRevResearch.2.033111} {\bibfield  {journal}
  {\bibinfo  {journal} {Phys. Rev. Research}\ }\textbf {\bibinfo {volume}
  {2}},\ \bibinfo {pages} {033111} (\bibinfo {year} {2020})}\BibitemShut
  {NoStop}%
\end{thebibliography}%
\clearpage
\newpage
\widetext
\appendix
\setcounter{figure}{0} 
\renewcommand{\thefigure}{S\arabic{figure}}
\setcounter{table}{0}
\renewcommand{\thetable}{S\arabic{figure}}
\begin{center}
\textbf{Supplemental Material:\\Connecting Finite-Temperature Dynamical and Excited-State Quantum Phase Transitions}
\end{center}

\begin{figure}[h]
\hspace{-3.4cm}\includegraphics[width=0.65\textwidth]{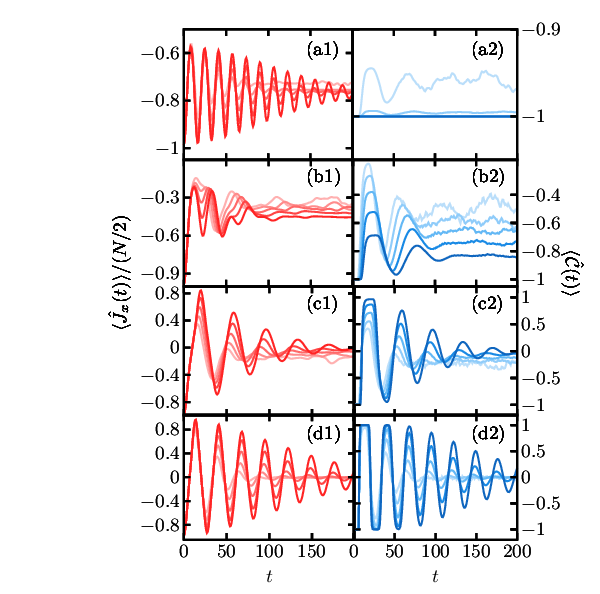}
\caption{Instantaneous values of the magnetization, $\hat{J}_{x}$, (left), and the $\hat{\mathcal{C}}$ operator, (right), following different quenches with $\beta=10$, $\lambda=0.5$ and $h_{i}=0$. (a1,2) $h_{f}=0.2$; (b1,2) $h_{f}=0.24$; (c1,2) $h_{f}=0.26$; (d1,2)  $h_{f}=0.3$. System sizes are $N=100,200,400,800,1600$ from light to dark color curves. The critical $h_{f}^{c}\approx 0.2464$.}
\label{figbeta10}
\end{figure}

\begin{figure}[h]
\hspace{-3.4cm}\includegraphics[width=0.65\textwidth]{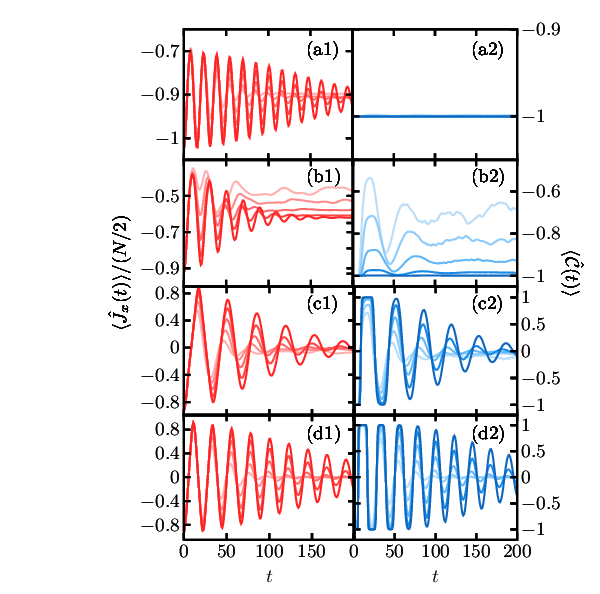}
\caption{Instantaneous values of the magnetization, $\hat{J}_{x}$, (left), and the $\hat{\mathcal{C}}$ operator, (right), following different quenches with $\beta=7$, $\lambda=0.5$ and $h_{i}=0$. (a1,2) $h_{f}=0.15$; (b1,2) $h_{f}=0.21$; (c1,2) $h_{f}=0.26$; (d1,2)  $h_{f}=0.33$. System sizes are $N=100,200,400,800,1600$ from light to dark color curves. The critical $h_{f}^{c}\approx 0.2311$. }
\label{figbeta7}
\end{figure}

\begin{figure}[h]
\hspace{-3.4cm}\includegraphics[width=0.65\textwidth]{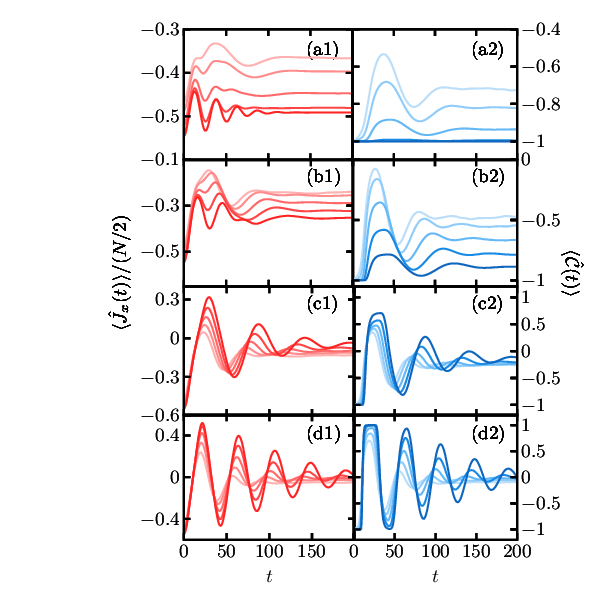}
\caption{Instantaneous values of the magnetization, $\hat{J}_{x}$, (left), and the $\hat{\mathcal{C}}$ operator, (right), following different quenches with $\beta=4.5$, $\lambda=0.5$ and $h_{i}=0$. (a1,2) $h_{f}=0.08$; (b1,2) $h_{f}=0.12$; (c1,2) $h_{f}=0.15$; (d1,2)  $h_{f}=0.18$. System sizes are $N=100,200,400,800,1600$ from light to dark color curves. The critical $h_{f}^{c}\approx 0.1378$. }
\label{figbeta4_5}
\end{figure}

\begin{figure}[h]
\hspace{-3.4cm}\includegraphics[width=0.65\textwidth]{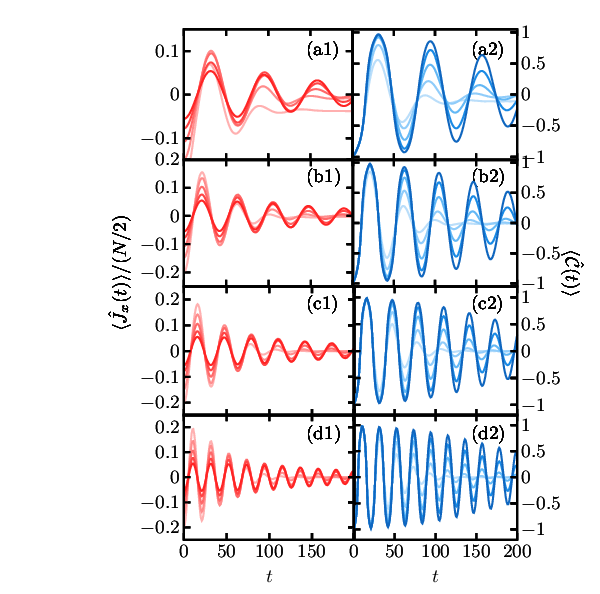}
\caption{Instantaneous values of the magnetization, $\hat{J}_{x}$, (left), and the $\hat{\mathcal{C}}$ operator, (right), following different quenches with $\beta=3.5$, $\lambda=0.5$ and $h_{i}=0$. (a1,2) $h_{f}=0.1$; (b1,2) $h_{f}=0.15$; (c1,2) $h_{f}=0.2$; (d1,2)  $h_{f}=0.3$. System sizes are $N=100,200,400,800,1600$ from light to dark color curves. Note that as $N$ increases, the oscillation in $\langle \hat{J}_{x}(t)\rangle$ gives an average value that approaches zero. Since our state is prepared with $h_{i}=0$, in the infinite-$N$ limit the magnetization should vanish at all times. Finite-size deviations from this limiting behavior are expected.}
\label{figbeta3_5}
\end{figure}

In this Supplemental Material we show the quench dynamics generated by initial states with inverse temperatures $\beta$ and $h_{f}$ different from those in the main text. The qualitative picture for $\beta=10,7,4.5$ in Figs.~\ref{figbeta10}, \ref{figbeta7}, and \ref{figbeta4_5} is essentially the same as in Fig.~\ref{fig:JxtCt} of the main text. As an exception to this behavior, in Fig.~\ref{figbeta3_5} we focus on $\beta=3.5<\beta_{c}$, for which the time evolution of the relevant observables always oscillates around zero irrespective of $h_{f}$. This is because for $\beta<\beta_{c}$, the most populated $j$-sector is always below the critical $j_{c}$, and therefore there is no ESQPT critical energy. As such, this does not contradict our conclusions, but rather reinforces them, because also for $\beta<\beta_c$ there is no DPT for quenches in $h$. Tables~\ref{tablebeta10}, \ref{tablebeta7}, and \ref{tablebeta4_5} compare the average energy of the quenched state with the range of ESQPT critical energies associated to the 95\% most populated $j$-sectors in the final Hamiltonian. Note that for the quench where $\beta<\beta_c$, Fig.~\ref{figbeta3_5} has no corresponding table as there is no ESQPT critical energy.

\begin{table}[h]
\begin{center}
 \begin{tabular}{||c c c||} 
 \hline
    Quench \hspace{0.1cm} & $\langle \varepsilon\rangle $ \hspace{0.1cm} & $[\varepsilon_{c,\min}, \varepsilon_{c,\max}]$\\ [0.5ex] 
 \hline\hline
$h_{f}=0.2$ \hspace{0.1cm} & $-0.2430\pm 0.0063$ \hspace{0.1cm} & $[-0.1988,-0.1955]$  \\[1ex]
 \hline
 $h_{f}=0.24$ \hspace{0.1cm} & $-0.2430\pm 0.0063$ \hspace{0.1cm} & $[-0.2385,-0.2346]$  \\[1ex]
 \hline

$h_{f}=0.26$ \hspace{0.1cm} & $-0.2430\pm 0.0063$ \hspace{0.1cm} & $[-0.2584,-0.2542]$  \\[1ex]
 \hline

 $h_{f}=0.3$ \hspace{0.1cm} & $-0.2430\pm 0.0063$ \hspace{0.1cm} & $[-0.2981,-0.2933]$  \\[1ex]
 \hline

\end{tabular}
\end{center}
\caption{Average energy, $\langle\varepsilon\rangle$, of the quenched state $h_{i}=0\to h_{f}$, for different values of $h_{f}$, and estimated maximal and minimal critical energies corresponding to the minimum and maximum $j$-sectors, respectively, with a cumulative probability of 95\%. System size is $N=1600$ and $\beta=10$. }
\label{tablebeta10}
\end{table}

\newpage

\begin{table}[h]
\begin{center}
 \begin{tabular}{||c c c||} 
 \hline
    Quench \hspace{0.1cm} & $\langle \varepsilon\rangle $ \hspace{0.1cm} & $[\varepsilon_{c,\min}, \varepsilon_{c,\max}]$\\ [0.5ex] 
 \hline\hline
$h_{f}=0.15$ \hspace{0.1cm} & $-0.2136\pm 0.0094$ \hspace{0.1cm} & $[-0.1418,-0.1354]$  \\[1ex]
 \hline
 $h_{f}=0.21$ \hspace{0.1cm} & $-0.2136\pm 0.0094$ \hspace{0.1cm} & $[-0.1985,-0.1895]$  \\[1ex]
 \hline

$h_{f}=0.26$ \hspace{0.1cm} & $-0.2136\pm 0.0094$ \hspace{0.1cm} & $[-0.2457,-0.2347]$  \\[1ex]
 \hline

 $h_{f}=0.33$ \hspace{0.1cm} & $-0.2136\pm 0.0094$ \hspace{0.1cm} & $[-0.3119,-0.2978]$  \\[1ex]
 \hline

\end{tabular}
\end{center}
\caption{Average energy, $\langle\varepsilon\rangle$, of the quenched state $h_{i}=0\to h_{f}$, for different values of $h_{f}$, and estimated maximal and minimal critical energies corresponding to the minimum and maximum $j$-sectors, respectively, with a cumulative probability of 95\%. System size is $N=1600$ and $\beta=7$. }
\label{tablebeta7}
\end{table}

\begin{table}[h]
\begin{center}
 \begin{tabular}{||c c c||} 
 \hline
    Quench \hspace{0.1cm} & $\langle \varepsilon\rangle $ \hspace{0.1cm} & $[\varepsilon_{c,\min}, \varepsilon_{c,\max}]$\\ [0.5ex] 
 \hline\hline
$h_{f}=0.08$ \hspace{0.1cm} & $-0.07522\pm 0.01108$ \hspace{0.1cm} & $[-0.0507,-0.0363]$  \\[1ex]
 \hline
 $h_{f}=0.12$ \hspace{0.1cm} & $-0.07522\pm 0.01108$ \hspace{0.1cm} & $[-0.07605,-0.0545]$  \\[1ex]
 \hline

$h_{f}=0.15$ \hspace{0.1cm} & $-0.07522\pm 0.01108$ \hspace{0.1cm} & $[-0.09506,-0.06806]$  \\[1ex]
 \hline

 $h_{f}=0.18$ \hspace{0.1cm} & $-0.07522\pm 0.01108$ \hspace{0.1cm} & $[-0.1141,-0.08168]$  \\[1ex]
 \hline

\end{tabular}
\end{center}
\caption{Average energy, $\langle\varepsilon\rangle$, of the quenched state $h_{i}=0\to h_{f}$, for different values of $h_{f}$, and estimated maximal and minimal critical energies corresponding to the minimum and maximum $j$-sectors, respectively, with a cumulative probability of 95\%. System size is $N=1600$ and $\beta=4.5$. }
\label{tablebeta4_5}
\end{table}

\newpage

\end{document}